\documentclass[onecolumn,12pt,journal,draftclsnofoot]{IEEEtran}
\usepackage[latin9]{inputenc}
\usepackage{geometry}
\usepackage{color}
\usepackage{float}
\usepackage{amsmath}
\usepackage{amssymb}
\usepackage{graphicx}

\makeatletter

\floatstyle{ruled}
\newfloat{algorithm}{tbp}{loa}
\providecommand{\algorithmname}{Algorithm}
\floatname{algorithm}{\protect\algorithmname}

\usepackage{amsfonts}
\usepackage{mathrsfs}
\usepackage{mathrsfs}\usepackage[noblocks]{authblk}
\usepackage[compress]{cite}
\usepackage{bm}
\usepackage{algorithm}
\usepackage{algorithmic}
\usepackage{epsfig}\usepackage{graphics}\usepackage{subfigure}\usepackage{epsfig}\usepackage{epstopdf}
\usepackage{graphics}\usepackage{subfigure}\usepackage{upgreek}\usepackage{caption}\usepackage{theorem}
\usepackage{breqn}
\usepackage{multicol}
\usepackage{eufrak}
\usepackage{eucal}
\usepackage{array}
\usepackage[compress]{cite}
\usepackage{algorithm}
\usepackage{algorithmic}
\allowdisplaybreaks[4]

\newtheorem{theorem}{Theorem}\newtheorem{lemma}{Lemma}\theoremheaderfont{\normalfont\bfseries}

\makeatother

\begin{document}
	\title{Stochastic Learning-Based Robust Beamforming Design for RIS-Aided
		Millimeter-Wave Systems in the Presence of Random Blockages}
	\author{Gui~Zhou, Cunhua~Pan, Hong~Ren, Kezhi~Wang, Maged Elkashlan, and
		Marco Di Renzo,~\IEEEmembership{Fellow, IEEE}
		\thanks{(Corresponding author: Cunhua Pan) G. Zhou, C. Pan and M. Elkashlan
			are with the School of Electronic Engineering and Computer Science
			at Queen Mary University of London, London E1 4NS, U.K. (e-mail: g.zhou,
			c.pan, maged.elkashlan@qmul.ac.uk). H. Ren is with the National Mobile
			Communications Research Laboratory, Southeast University, Nanjing
			210096, China. (hren@seu.edu.cn). K. Wang is with Department of Computer
			and Information Sciences, Northumbria University, UK. (e-mail: kezhi.wang@northumbria.ac.uk). M. Di Renzo is with Universit\'e Paris-Saclay, CNRS, CentraleSup\'elec, Laboratoire des Signaux et Syst\`emes, 3 Rue Joliot-Curie, 91192 Gif-sur-Yvette, France. (marco.di-renzo@universite-paris-saclay.fr) M. Di Renzo's work was supported in part by the European Commission through the H2020 ARIADNE project under grant agreement 675806.}
	}

\maketitle
\begin{abstract}
	A fundamental challenge for millimeter wave (mmWave) communications
	lies in its sensitivity to the presence of blockages, which impact
	the connectivity of the communication links and ultimately the reliability
	of the network. In this paper, we analyze a mmWave communication system
	assisted by multiple reconfigurable intelligent surface (RISs) for
	enhancing the network reliability and connectivity in the presence
	of random blockages. To enhance the robustness of beamforming in the
	presence of random blockages, we formulate a stochastic optimization
	problem based on the minimization of the sum outage probability. To
	tackle the proposed optimization problem, we introduce a low-complexity
	algorithm based on the stochastic block gradient descent method, which
	learns sensible blockage patterns without searching for all combinations
	of potentially blocked links. Numerical results confirm the performance
	benefits of the proposed algorithm in terms of outage probability
	and effective data rate. 
\end{abstract}

\section{Introduction}

Due to the abundance of available frequency bandwidth, millimeter
wave (mmWave) communication is envisioned to be a potential technology
to meet the high data rate demand of current wireless networks. In
addition, due to the small wavelength, a large number of antenna elements
can be packed in antenna arrays of reasonable size, which can compensate
for the severe path loss caused by the high transmission frequency
and can mitigate the inter-user interference by capitalizing on the
design of high-directional beams \cite{mmWave2013}. Moreover, the
use of hybrid analog-digital array reduces the cost and power consumption
of using many radio frequency (RF) chains and full-digital signal
processing baseband units at mmWave base stations (BSs) \cite{Boya-1}.

However, the mmWave signals experience high penetration losses and
a low diffraction from objects \cite{blockage2019}, which make mmWave
systems highly sensible to the presence of spatial blockages (e.g.,
buildings, human beings, etc.) and degrades the reliability of the
communication links. To
address this challenge, recent research works \cite{blockage-robust2,blockage-robust3}
have proposed some robust beamforming designs to tackle the channel
uncertainties due to the presence of random blockages. In particular,
by capitalizing on the predicted blockage probability, \cite{blockage-robust2}
proposed a worst-case robust coordinated multipoint (CoMP) beamforming
design by considering all possible combinations of blockage patterns.
To reduce the complexity and improve the robustness of mmWave communication
that suffers from the presence of random blockages, an outage-minimum
strategy based on a stochastic optimization method was proposed in
\cite{blockage-robust3}.

However, the above methods are only suitable for CoMP scenarios,
where the space diversity gain from the multiple BSs reduces
the outage caused by the presence of random blockages.
Deploying multiple BSs, however, increases the hardware cost and power
consumption. To overcome these challenges, the emerging technology
of reconfigurable intelligent surfaces (RISs) was recently proposed
as a promising solution for establishing alternative communication
routes at a low cost, high energy efficiency, and high reliability
\cite{Marco-3,Marco-4,Marco-5,Marco-6,Boya-2,Boya-3}. An RIS is a
surface made of nearly-passive and reconfigurable scatterers,
which are capable of modifying the incident radio waves, so as to
enhance the received signals at some specified locations \cite{Pan2019intelleget,Pan2019multicell,Baitong2019,Marco11,Gui-letter,Marco12,Boya-1}.

In this paper, motivated by these considerations, we study an mmWave
RISs-aided communication system, and optimize the hybrid analog-digital
beamforming at the BS and the passive beamforming at the RISs by explicitly
taking into account the presence of random blockages. To enhance the
robustness of the considered system, we study a sum-outage probability
minimization problem that minimizes the rate at which the system quality
of service (QoS), which is quantified in terms of outage probability,
is not fulfilled. The formulated stochastic optimization problem with
coupled variables is solved by using the block stochastic gradient
descent (BSGD) method. The obtained simulation results show the performance
benefits of the proposed scheme in terms of outage probability and
effective data rate.

\section{System Model}

\begin{figure}
	\centering \includegraphics[width=2.8in,height=1.6in]{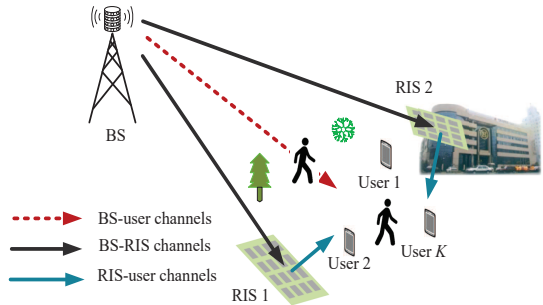}
	\caption{RISs-aided mmWave communication system.}
	\label{system-model} 
\end{figure}

\subsection{Signal Model}

As shown in Fig. \ref{system-model}, we consider the downlink of
an mmWave system in which a BS equipped with a uniform linear array
(ULA) with $N$ antennas and $N_{RF}$ RF chains serves $K$ single-antenna
users (denoted by $\mathcal{K}\triangleq\{1,...,K\}$) in the presence
of random blockages, where $K\leq N_{RF}\ll N$. We assume that $U$
RISs deployed on e.g., the facade of some buildings, have the capability
of passively reflecting the signals transmitted from the BS to the
users. Each RIS is made of $M$ passive reflecting elements that are
arranged in a uniform planar array (UPA). It is assumed that the phase
shifts of the RIS are computed by the BS and are then sent to the
RIS controller through dedicated control channels \cite{Pan2019intelleget,Pan2019multicell}.
The BS adopts a hybrid precoding architecture, in which each RF chain
is connected to all the available antennas, and transmits Gaussian
data symbols $\mathbf{s}=[s_{1},\cdots,s_{K}]^{\mathrm{T}}\in\mathbb{C}^{K\times1}$
to the users through a digital precoding matrix $\mathbf{D}=[{\bf \mathbf{d}}_{1},\cdots,{\bf \mathbf{d}}_{K}]\in\mathbb{C}^{N_{RF}\times K}$
and an analog precoding matrix $\mathbf{A}\in\mathbb{C}^{N\times N_{RF}}$.
The transmit power of the BS fulfills the constraint $||\mathbf{A}\mathbf{D}||_{F}^{2}\leq P_{max}$,
where $P_{max}$ is the total transmit power limit. Each entry of
$\mathbf{A}$ is constrained to have a unit modulus, i.e., $\mathbf{A}\in\mathcal{S}_{A}$
where $\mathcal{S}_{A}\triangleq\{\mathbf{A}||[\mathbf{A}]_{m,n}|^{2}=1,\forall m,n\}$,
and the symbol $[\cdot]_{m,n}$ denotes the $(m,n)$-th element of
a matrix. The received signal intended to the $k$-th user can be
formulated as 
\begin{align}
y_{k} & =(\mathbf{h}_{\mathrm{b},k}^{\mathrm{H}}+\sum_{u=1}^{U}\mathbf{h}_{\mathrm{i},uk}^{\mathrm{H}}\mathbf{E}_{u}\mathbf{H}_{\mathrm{bi},u})\mathbf{A}\mathbf{D}{\bf s}+n_{k}\nonumber \\
& =(\mathbf{h}_{\mathrm{b},k}^{\mathrm{H}}+\mathbf{h}_{\mathrm{i},k}^{\mathrm{H}}\mathbf{E}\mathbf{H}_{\mathrm{bi}})\mathbf{A}\mathbf{D}{\bf s}+n_{k},\label{eq:1}
\end{align}
where $\mathbf{E}_{u}=\mathrm{diag}([e_{(u-1)M+1},\ldots,e_{uM}])$
is the reflection coefficient matrix (also known as the passive beamforming
matrix) of the $u$-th RIS and $n_{k}\sim\mathcal{CN}(0,\sigma_{k}^{2})$
is the additive white Gaussian noise (AWGN). The channels of the BS-user
$k$, BS-RIS $u$, and RIS $u$-user $k$ links are denoted by $\mathbf{h}_{\mathrm{b},k}\in\mathbb{C}^{N\times1}$,
$\mathbf{H}_{\mathrm{bi},u}\in\mathbb{C}^{M\times N}$ and $\mathbf{h}_{\mathrm{i},uk}\in\mathbb{C}^{M\times1}$,
respectively. With this notation, the matrices in (\ref{eq:1}) are
defined as $\mathbf{H}_{\mathrm{bi}}=[\mathbf{H}_{\mathrm{bi},1}^{\mathrm{H}},...,\mathbf{H}_{\mathrm{bi},U}^{\mathrm{H}}]^{\mathrm{H}}$
$\mathbf{h}_{\mathrm{i},k}^{\mathrm{H}}=[\mathbf{h}_{\mathrm{i},1k}^{\mathrm{H}},...,\mathbf{h}_{\mathrm{i},Uk}^{\mathrm{H}}]^{\mathrm{H}}$
and $\mathbf{E}=\mathrm{diag}(\mathbf{E}_{1},...,\mathbf{E}_{U})$.

Furthermore, denoting by $\mathbf{H}_{k}=\left[\begin{array}{c}
\mathrm{diag}(\mathbf{h}_{\mathrm{i},k}^{\mathrm{H}})\mathbf{H_{\mathrm{bi}}}\\
\mathbf{h}_{\mathrm{b},k}^{\mathrm{H}}
\end{array}\right]\in\mathbb{C}^{(M+1)\times N}$ the equivalent channel of the BS to the $k$-th user and by $\mathbf{e}=[e_{1},\ldots e_{UM},1]^{\mathrm{T}}\in\mathbb{C}^{(UM+1)\times1}$
the equivalent reflection coefficient vector that belongs to the set
$\mathcal{S}_{e}=\{\mathbf{e}\mid|e_{m}|^{2}=1,1\leq m\leq UM,e_{UM+1}=1\}$,
(\ref{eq:1}) can be rewritten as $y_{k}=\mathbf{e}^{\mathrm{H}}\mathbf{H}_{k}\mathbf{A}\mathbf{D}{\bf s}+n_{k},\forall k\in\mathcal{K},$
and the corresponding achievable signal-to-interference-plus-noise
ratio (SINR), $\Omega_{k}\left(\mathbf{D},\mathbf{A},\mathbf{e}\right)$,
can be written as 
\begin{equation}
\Omega_{k}\left(\mathbf{D},\mathbf{A},\mathbf{e}\right)=\frac{|\mathbf{e}^{\mathrm{H}}\mathbf{H}_{k}\mathbf{A}{\bf d}_{k}|^{2}}{\sum_{i\neq k}^{K}|\mathbf{e}^{\mathrm{H}}\mathbf{H}_{k}\mathbf{A}{\bf d}_{i}|^{2}+\sigma_{k}^{2}}.\label{eq:Rate-k-1}
\end{equation}

\subsection{Channel Model}

Based on \cite{mmWave-channel}, a geometric channel model can be used to characterize the mmWave channel.  Assume that there are $L_{BU}$, $L_{BI}$ and
$L_{IU}$ propagation paths that characterize the BS-user links, the
BS-RIS links and the RIS-user links respectively, then we have 
\begin{align}
\mathbf{h}_{\mathrm{b},k} & =\sqrt{\frac{1}{L_{BU}}}\sum_{l=1}^{L_{BU}}g_{k,l}^{\mathrm{b}}\mathbf{a}_{L}\left(\theta_{k,l}^{\mathrm{b},t}\right),\forall k\in\mathcal{K},\label{eq:3}\\
\mathbf{h}_{\mathrm{i},k} & =\sqrt{\frac{1}{L_{IU}}}\sum_{l=1}^{L_{IU}}g_{k,l}^{\mathrm{i}}\mathbf{a}_{P}\left(\theta_{k,l}^{\mathrm{\mathrm{i}},t},\phi_{k,l}^{\mathrm{i},t}\right),\forall k\in\mathcal{K},\label{eq:4}\\
\mathbf{H_{\mathrm{bi}}} & =\sqrt{\frac{1}{L_{BI}}}\sum_{l=1}^{L_{BI}}g_{l}^{\mathrm{\mathrm{bi}}}\mathbf{a}_{P}\left(\theta_{l}^{\mathrm{\mathrm{i}},r},\phi_{l}^{\mathrm{\mathrm{i}},r}\right)\mathbf{a}_{L}\left(\theta_{l}^{\mathrm{b},t}\right)^{\mathrm{H}},\label{eq:5}
\end{align}
where $\{g_{k,l}^{\mathrm{b}},g_{k,l}^{\mathrm{i}},g_{l}^{\mathrm{\mathrm{bi}}}\}$
denote the large-scale fading coefficients. Define $g\in\{g_{k,l}^{\mathrm{b}}\textrm{ for }\forall k,g_{k,l}^{\mathrm{i}}\textrm{ for }\forall k,g_{l}^{\mathrm{\mathrm{bi}}}\}$,
then $g$ has distribution $\mathcal{CN}(0,10^{\frac{\mathrm{PL}}{10}})$,
where $\mathrm{PL}=-C_{0}-10\alpha\log_{10}(D)-\zeta$ dB, $C_{0}$
is the path loss at a reference distance of one meter, $D$ is the
link distance (in meters), $\alpha$ is the pathloss exponent and
$\zeta$ is the lognormal shadowing \cite{mmWave-channel}. Also,
$\mathbf{a}_{L}\left(\theta\right)$ and $\mathbf{a}_{P}\left(\theta,\phi\right)$
are the steering vectors of the ULA and UPA, respectively.

According to \cite{blockage2019,blockage-robust2,blockage-robust3},
the BS-user links may be obstructed by the presence of a random blockage
with a certain probability, while the RIS-related links can be assumed
not to be affected by blockages, since the locations of the RISs can
be appropriately optimized in order to ensure line of sight transmission.
Due to the fact that the communication links in the mmWave frequency
band are severely attenuated by the presence of blockages, the achievable
data rate may be significantly reduced. In order to investigate the
impact of the channel uncertainties caused by the presence of random
blockages, we adopt a recently proposed probabilistic model for the
BS-user links \cite{blockage-robust3}. In particular, the channels
between the BS and the users are modeled as 
\begin{equation}
\mathbf{h}_{\mathrm{b},k}=\sqrt{\frac{1}{L_{BU}}}\sum_{l=1}^{L_{BU}}\gamma_{k,l}g_{k,l}^{\mathrm{b}}\mathbf{a}_{L}\left(\theta_{k,l}^{\mathrm{b},t}\right),\forall k\in\mathcal{K},\label{eq:5-1}
\end{equation}
where the random variable $\gamma_{k,l}\in\{0,1\}$ is a blockage
parameter that is distributed according to a Bernoulli distribution.
In particular, the corresponding blockage probability is denoted by
$p_{k,l}$. In this work, we assume that the blockage probability
is known at the BS and that it is used for robust beamforming design.

\subsection{Problem Formulation}

In the presence of random blockages, we aim to propose a robust mmWave
beamforming scheme with the objective of minimizing the sum outage
probability \cite{blockage-robust3}. The formulated optimization
problem is given by\begin{subequations}\label{Pro:min-power} 
	\begin{align}
	\mathop{\min}\limits _{\mathbf{D},\mathbf{A},\mathbf{e}} & \;\;\sum_{k\in\mathcal{K}}\mathrm{Pr}\{\Omega_{k}\left(\mathbf{D},\mathbf{A},\mathbf{e}\right)\leq\omega_{k}\}\label{eq:P-obj}\\
	\textrm{s.t.} & \thinspace\thinspace\thinspace||\mathbf{A}\mathbf{D}||_{F}^{2}\leq P_{max}\label{eq:p-c1-1}\\
	& \thinspace\thinspace\thinspace\mathbf{A}\in\mathcal{S}_{A}\label{eq:p-c1-2}\\
	& \thinspace\thinspace\thinspace\mathbf{e}\in\mathcal{S}_{e},\label{eq:p-c1-3}
	\end{align}
\end{subequations}where $\omega_{k}>0$ is the SINR reliability threshold
of the $k$-th user and $\mathrm{Pr}\{\Omega_{k}\left(\mathbf{D},\mathbf{A},\mathbf{e}\right)\leq\omega_{k}\}$
denotes the probability that the required SINR cannot be satisfied.

In contrast to the traditional deterministic formulation which might
not always have feasible solutions due to the QoS constraints, Problem
(\ref{Pro:min-power}) has always a feasible solution that ensures
the desired QoS target minimum outage probability.

\section{Beamforming design}

Problem (\ref{Pro:min-power}) is challenging to solve due to the
absence of a closed-form expression for the objective function in
(\ref{eq:P-obj}), the non-convex unit-modulus constraints in (\ref{eq:p-c1-2})
and (\ref{eq:p-c1-3}), and that fact that the optimization variables
are tightly coupled. In the following, we propose a robust beamforming
design algorithm under the stochastic-learning-based alternating optimization
(AO) framework.

\subsection{Problem Transformation}

To start with, we rewrite $\mathrm{Pr}\{\Omega_{k}\left(\mathbf{D},\mathbf{A},\mathbf{e}\right)\leq\omega_{k}\}$
as $\mathbb{E}_{\mathbf{H}_{k}}[\mathbb{I}_{\Omega_{k}\leq\omega_{k}}]$,
where $\mathbb{I}_{\Omega_{k}\leq\omega_{k}}$ is the step function.
Since the step function is non-differentiable, we approximate it with
the following smooth hinge surrogate function \cite{hing2008} 
\[
u_{k}\left(\mathbf{X}\right)=\begin{cases}
0 & \textrm{if }1-\frac{\Omega_{k}\left(\mathbf{X}\right)}{\omega_{k}}<0\\
\frac{1}{2\epsilon}\left(1-\frac{\Omega_{k}\left(\mathbf{X}\right)}{\omega_{k}}\right)^{2} & \textrm{otherwise}\\
1-\frac{\Omega_{k}\left(\mathbf{X}\right)}{\omega_{k}}-\frac{\epsilon}{2} & \textrm{if }1-\frac{\Omega_{k}\left(\mathbf{X}\right)}{\omega_{k}}>\epsilon,
\end{cases}
\]
where $0<\epsilon\ll1$ and $\mathbf{X}\triangleq\{\mathbf{D},\mathbf{A},\mathbf{e}\}$
is a short-hand notation that collects of all the variables of interest.
By replacing the step function $\mathbb{I}_{\Omega_{k}\leq\omega_{k}}$
with its smooth approximation $u_{k}\left(\mathbf{X}\right)$, we
obtain the following approximated reformulation for Problem (\ref{Pro:min-power})
\begin{subequations}\label{Pro:min-power-2-1} 
	\begin{align}
	\mathop{\min}\limits _{\mathbf{X}} & \;\;\sum_{k\in\mathcal{K}}\mathbb{E}_{\mathbf{H}_{k}}[u_{k}\left(\mathbf{X}\right)]\label{eq:P-obj-2}\\
	\textrm{s.t.} & \thinspace\thinspace\thinspace(\ref{eq:p-c1-1}),(\ref{eq:p-c1-2}),(\ref{eq:p-c1-3}).
	\end{align}
\end{subequations}

Problem (\ref{Pro:min-power-2-1}) can be viewed as a risk minimization
problem \cite{SGD2020MAG}, which has been studied thoroughly in several
application areas such as wireless resource optimization, compressive
sensing, machine learning, etc. It can be solved efficiently by using
stochastic optimization methods, which are widely adopted due to their
easy-to-implement features.

Since the BS is assumed to know the channel state information (CSI),
namely the channel gain and the AoDs, the expectation in (\ref{eq:P-obj-2})
is computed with respect to  $\gamma_{k,l}$
in (\ref{eq:5-1}). By assuming that the blockage probability can
be predicted with good accuracy, we can generate $\gamma_{k,l}$ randomly
and can construct the training data sample set $\mathcal{H}=\{\mathbf{H}_{k}^{(t)},\forall k\in\mathcal{K}\}_{t=1}^{T}$
that is utilized for stochastic optimization. Accordingly, an appropriate
surrogate for the risk function $\mathbb{E}_{\mathbf{H}_{k}}[u_{k}\left(\mathbf{X}\right)]$
is often assumed to be $\frac{1}{T}\sum_{t=1}^{T}u_{k}(\mathbf{X},\mathbf{H}_{k}^{(t)})$,
which is usually referred to as the empirical risk function \cite{SGD2020MAG}.
Therefore, the resulting empirical risk minimization (ERM) problem
is\begin{subequations}\label{Pro:min-power-2-1-1}
	\begin{align}
	\mathop{\min}\limits _{\mathbf{X}} & \;\;\frac{1}{T}\sum_{t=1}^{T}\sum_{k\in\mathcal{K}}u_{k}(\mathbf{X},\mathbf{H}_{k}^{(t)})\label{eq:P-obj-2-1}\\
	\textrm{s.t.} & \thinspace\thinspace\thinspace(\ref{eq:p-c1-1}),(\ref{eq:p-c1-2}),(\ref{eq:p-c1-3}).
	\end{align}
\end{subequations}

A popular method to solve ERM problems is the stochastic gradient
descent (SGD) \cite{SGD1998}, which we leverage to solve Problem
(\ref{Pro:min-power-2-1-1}) by alternately optimizing one of the
block variables $\{\mathbf{D},\mathbf{A},\mathbf{e}\}$ while keeping
the others fixed. This algorithm is usually referred to as the block
stochastic gradient descent (BSGD) method \cite{BSDGD}.

\subsection{Algorithm Description}

To introduce the proposed algorithm, we denote by $\mathcal{S}_{D}\triangleq\{\mathbf{D}|\thinspace\thinspace||\mathbf{A}\mathbf{D}||_{F}^{2}\leq P_{max}\}$
the set of $\mathbf{D}$ with fixed $\mathbf{A}$ and by $\mathcal{P}_{\mathcal{S}_{z}}({\bf z})$
the Euclidean projection from a point ${\bf z}$ onto a set $\mathcal{S}_{z}$,
i.e., $\mathcal{P}_{\mathcal{S}_{z}}({\bf z})=\mathrm{arg}\min_{{\bf y}\in\mathcal{S}_{z}}||{\bf y}-{\bf z}||$.
\begin{algorithm}
	\caption{BSGD-OutMin Algorithm}
	\label{Algorithm-1} \begin{algorithmic}[1] \REQUIRE Initialize
		$\mathbf{D}^{(0)}$, $\mathbf{A}^{(0)}$, $\mathbf{e}^{(0)}$, and
		the data set $\mathcal{H}$. Set $t=1$ and $T_{max}=10^{5}.$
		
		\REPEAT
		
		\STATE Sample the data $\mathbf{H}_{k}^{(t)}\thinspace\thinspace\thinspace\forall k\in\mathcal{K}$
		from $\mathcal{H}$.
		
		\STATE ${\bf D}^{(t)}=\mathcal{P}_{\mathcal{S}_{D}}({\bf D}^{(t-1)}-\alpha_{t}\sum_{k\in\mathcal{K}}\nabla_{\mathbf{D}}u_{k}(\mathbf{X},\mathbf{H}_{k}^{(t)})).$
		
		\STATE ${\bf A}^{(t)}=\mathcal{P}_{\mathcal{S}_{A}}({\bf A}^{(t-1)}-\alpha_{t}\sum_{k\in\mathcal{K}}\nabla_{\mathbf{A}}u_{k}(\mathbf{X},\mathbf{H}_{k}^{(t)})).$
		
		\STATE ${\bf e}^{(t)}=\mathcal{P}_{\mathcal{S}_{e}}({\bf e}^{(t-1)}-\alpha_{t}\sum_{k\in\mathcal{K}}\nabla_{\mathbf{e}}u_{k}(\mathbf{X},\mathbf{H}_{k}^{(t)})).$
		
		\STATE $t=t+1$.
		
		\UNTIL The objective value in (\ref{eq:P-obj-2-1}) converges.
		
	\end{algorithmic} 
\end{algorithm}

Algorithm \ref{Algorithm-1} summarizes the proposed BSGD-based outage
minimum robust hybrid beamforming design for RIS-aided mmWave systems
in which the BS-user links undergo random blockages. The proposed
algorithm is referred to as BSGD-OutMin. It is not difficult to carry
out the projection operations in Algorithm \ref{Algorithm-1} as follows
\begin{align*}
\mathcal{P}_{\mathcal{S}_{D}}({\bf Z}) & =\frac{\left(\mathbf{Z}\right)}{||\mathbf{A}^{(t-1)}\mathbf{Z}||_{F}}\sqrt{P_{max}},\\
\mathcal{P}_{\mathcal{S}_{A}}({\bf Z}) & =\exp\left\{ j\angle\mathbf{Z}\right\} ,\\
\mathcal{P}_{\mathcal{S}_{e}}({\bf z}) & =\exp\left\{ j\angle\left(\mathbf{z}/\left[\mathbf{z}\right]_{UM+1}\right)\right\} .
\end{align*}

\paragraph{Gradient of $u_{k}(\mathbf{X},\mathbf{H}_{k}^{(t)})$}

Define $\nabla_{\mathbf{X}}u_{k}(\mathbf{X},\mathbf{H}_{k}^{(t)}))$
the stochastic gradients in Algorithm \ref{Algorithm-1}. It is worth
noting that the normalization factor $1/N$ in $\sum_{k\in\mathcal{K}}\nabla_{\mathbf{X}}u_{k}(\mathbf{X},\mathbf{H}_{k}^{(t)}))$
is omitted, since the stochastic gradient $\nabla_{\mathbf{X}}u_{k}(\mathbf{X},\mathbf{H}_{k}^{(t)})$
is an unbiased estimator for the batch gradient $\frac{1}{T}\sum_{t=1}^{T}\nabla_{\mathbf{X}}u_{k}(\mathbf{X},\mathbf{H}_{k}^{(t)})$
\cite{SGD2020MAG}, i.e., $\mathbb{E}_{\mathbf{H}_{k}^{(t)}}[\nabla_{\mathbf{X}}u_{k}(\mathbf{X},\mathbf{H}_{k}^{(t)})|\mathbf{x}^{(t-1)}]=\frac{1}{T}\sum_{t=1}^{T}\nabla_{\mathbf{X}}u_{k}(\mathbf{X},\mathbf{H}_{k}^{(t)})$.

The partial gradient of $u_{k}\left(\mathbf{X},\mathbf{H}_{k}^{(t)}\right)$
with respect to $\mathbf{x}$ is 
\begin{align}
\nabla_{\mathbf{x}}u_{k}(\mathbf{X},\mathbf{H}_{k}^{(t)})=\begin{cases}
0 & \textrm{if }1-\frac{\Omega_{k}\left(\mathbf{X},\mathbf{H}_{k}^{(t)}\right)}{\omega_{k}}<0\\
\frac{\frac{\Omega_{k}\left(\mathbf{X},\mathbf{H}_{k}^{(t)}\right)}{\omega_{k}}-1}{\epsilon}\frac{\nabla_{\mathbf{x}}\Omega_{k}\left(\mathbf{X},\mathbf{H}_{k}^{(t)}\right)}{\omega_{k}} & \textrm{otherwise}\\
-\frac{\nabla_{\mathbf{x}}\Omega_{k}\left(\mathbf{X},\mathbf{H}_{k}^{(t)}\right)}{\omega_{k}} & \textrm{if }1-\frac{\Omega_{k}\left(\mathbf{X},\mathbf{H}_{k}^{(t)}\right)}{\omega_{k}}>\epsilon.
\end{cases}\label{eq:grad-x}
\end{align}

In order to compute the partial gradient $\nabla\Omega_{k}\left(\mathbf{X},\mathbf{H}_{k}^{(t)}\right)$
with respect to any $\mathbf{X}\in\{\mathbf{D},\mathbf{A},\mathbf{e}\}$,
it is necessary to apply some mathematical transformations to the
SINR in (\ref{eq:Rate-k-1}). In particular, as far as $\mathbf{D}$
is concerned, (\ref{eq:Rate-k-1}) can be rewritten as 
\begin{equation}
\Omega_{k}\left(\mathbf{X},\mathbf{H}_{k}^{(t)}\right)=\frac{\mathrm{vec}({\bf D})^{\mathrm{H}}\mathbf{Q}_{\mathbf{D},k}^{(t)}\mathrm{vec}({\bf D})}{\mathrm{vec}({\bf D})^{\mathrm{H}}\overline{\mathbf{Q}}_{\mathbf{D},k}^{(t)}\mathrm{vec}({\bf D})+\sigma_{k}^{2}},\label{eq:d-1}
\end{equation}
where $\mathbf{Q}_{\mathbf{D},k}^{(t)}=\textrm{diag}(\mathbf{i}_{k})\otimes\mathbf{A}^{\mathrm{H}}\mathbf{H}_{k}^{(t),\mathrm{H}}\mathbf{e}\mathbf{e}^{\mathrm{H}}\mathbf{H}_{k}^{(t)}\mathbf{A}$,
$\overline{\mathbf{Q}}_{\mathbf{D},k}^{(t)}=\textrm{diag}(\mathbf{\overline{i}}_{k})\otimes\mathbf{A}^{\mathrm{H}}\mathbf{H}_{k}^{(t),\mathrm{H}}\mathbf{e}\mathbf{e}^{\mathrm{H}}\mathbf{H}_{k}^{(t)}\mathbf{A}$,
and $\mathbf{i}_{k}$ is the $k$-th column of the $K\times K$ identity
matrix $\mathbf{I}_{K}$, $\mathbf{\overline{i}}_{k}$ denotes the
one's complement of $\mathbf{i}_{k}$, and $\otimes$ is the Kronecker
product.

As far as $\mathbf{A}$ is concerned, (\ref{eq:Rate-k-1}) is equivalent
to 
\begin{equation}
\Omega_{k}\left(\mathbf{X},\mathbf{H}_{k}^{(t)}\right)=\frac{\mathrm{vec}({\bf A})^{\mathrm{H}}\mathbf{Q}_{\mathbf{A},k}^{(t)}\mathrm{vec}({\bf A})}{\mathrm{vec}({\bf A})^{\mathrm{H}}\overline{\mathbf{Q}}_{\mathbf{A},k}^{(t)}\mathrm{vec}({\bf A})+\sigma_{k}^{2}},\label{eq:a1}
\end{equation}
where $\mathbf{Q}_{\mathbf{A},k}^{(t)}=\mathbf{d}_{k}^{*}\mathbf{d}_{k}^{\mathrm{T}}\otimes\mathbf{H}_{k}^{(t),\mathrm{H}}\mathbf{e}\mathbf{e}^{\mathrm{H}}\mathbf{H}_{k}^{(t)}$
and $\overline{\mathbf{Q}}_{\mathbf{A},k}^{(t)}=\mathbf{D}_{-k}^{*}\mathbf{D}_{-k}^{\mathrm{T}}\otimes\mathbf{H}_{k}^{(t),\mathrm{H}}\mathbf{e}\mathbf{e}^{\mathrm{H}}\mathbf{H}_{k}^{(t)}$.

As far as $\mathbf{e}$ is concerned, (\ref{eq:Rate-k-1}) is equivalent
to 
\begin{equation}
\Omega_{k}\left(\mathbf{X},\mathbf{H}_{k}^{(t)}\right)=\frac{\mathbf{e}^{\mathrm{H}}\mathbf{Q}_{\mathbf{e},k}^{(t)}\mathbf{e}}{\mathbf{e}^{\mathrm{H}}\overline{\mathbf{Q}}_{\mathbf{e},k}^{(t)}\mathbf{e}+\sigma_{k}^{2}},\label{eq:e1}
\end{equation}
where $\mathbf{Q}_{\mathbf{e},k}^{(t)}=\mathbf{H}_{k}^{(t)}\mathbf{A}{\bf d}_{k}{\bf d}_{k}^{\mathrm{H}}\mathbf{A}^{\mathrm{H}}\mathbf{H}_{k}^{(t),\mathrm{H}}$
and $\overline{\mathbf{Q}}_{\mathbf{e},k}^{(t)}=\mathbf{H}_{k}^{(t)}\mathbf{A}\sum_{i\neq k}^{K}{\bf d}_{i}{\bf d}_{i}^{\mathrm{H}}\mathbf{A}^{\mathrm{H}}\mathbf{H}_{k}^{(t),\mathrm{H}}$.

Equations (\ref{eq:d-1})-(\ref{eq:e1}) have the same mathematical
structure. Define $\mathbf{x}\in\{\mathrm{vec}({\bf D}),\mathrm{vec}({\bf A}),\mathbf{e}\}$,
then the unified partial gradient for any $\mathbf{x}$ is given by
\[
\nabla_{\mathbf{x}}\Omega_{k}\left(\mathbf{x},\mathbf{H}_{k}^{(t)}\right)=\frac{\mathbf{Q}_{\mathbf{x},k}^{(t)}\mathbf{x}}{v_{k}}-\frac{\mathbf{x}^{\mathrm{H}}\mathbf{Q}_{\mathbf{x},k}^{(t)}\mathbf{x}}{v_{k}^{2}}\overline{\mathbf{Q}}_{\mathbf{x},k}^{(t)}\mathbf{x},
\]
where $v_{k}=\mathbf{x}^{\mathrm{H}}\overline{\mathbf{Q}}_{\mathbf{x},k}^{(t)}\mathbf{x}+\sigma_{k}^{2}$.

\paragraph{Initial point}

Problem (\ref{Pro:min-power}) has multiple local minima
points due to the non-convex constraints $\mathbf{A}\in\mathcal{S}_{A}$
and $\mathbf{e}\in\mathcal{S}_{e}$. The accurate selection of the
initial points in Algorithm \ref{Algorithm-1} plays an important
role for the convergence speed and the optimality of the obtained
local solution. To that end, we first initialize $\mathbf{e}$ to
maximize the equivalent total channel gain, which results in the following
optimization problem \begin{subequations}\label{Pro:initiali-e}
	\begin{align}
	\mathop{\max}\limits _{\mathbf{e}\in\mathcal{S}_{e}} & \;\;\sum_{k\in\mathcal{K}}||\mathbf{e}^{\mathrm{H}}\mathbf{H}_{k}^{(0)}||_{2}^{2},\label{eq:d}
	\end{align}
\end{subequations}where $\mathbf{H}_{k}^{(0)}\thinspace\thinspace\forall k$
are the deterministic matrices with blockage probability $p_{k,l}=0,\forall k,l$,
i.e., $\gamma_{k,l}=1,\forall k,l$.

The objective function in (\ref{Pro:initiali-e}) is convex and can
be linearized by using its first-order Taylor approximation, i.e.,
$2\textrm{\ensuremath{\mathrm{Re}}}\{\mathbf{e}^{[n],\mathrm{H}}\mathbf{H}^{(0)}\mathbf{H}^{(0),\mathrm{H}}\mathbf{e}\}-\mathbf{e}^{[n],\mathrm{H}}\mathbf{H}^{(0)}\mathbf{H}^{(0),\mathrm{H}}\mathbf{e}^{[n]}$,
where $\mathbf{H}^{(0)}=[\mathbf{H}_{1}^{(0)},\ldots,\mathbf{H}_{K}^{(0)}]$
and $\mathbf{e}^{[n]}$ is the optimal solution obtained at the $n$-th
iteration, and $\textrm{\ensuremath{\mathrm{Re}}}\{\cdot\}$ denotes
the real part of a complex number. Therefore, the optimal solution
to Problem (\ref{Pro:initiali-e}) at the $(n+1)$-th iteration is
\[
\mathbf{e}^{[n+1]}=\exp\left\{ j\angle\left(\mathbf{H}^{(0)}\mathbf{H}^{(0),\mathrm{H}}\mathbf{e}^{[n]}/\left[\mathbf{H}^{(0)}\mathbf{H}^{(0),\mathrm{H}}\mathbf{e}^{[n]}\right]_{M+1}\right)\right\} .
\]
Furthermore, $\mathbf{A}$ is initialized to align the phases of the
equivalent channel to the mmWave BS, i.e., 
\[
\mathbf{A}^{(0)}=\exp\left\{ j\angle\left(\mathbf{H}^{(0)}(\mathbf{I}_{K}\otimes\mathbf{e}^{(0)})\right)\right\} .
\]

\paragraph{Convergence analysis}

It is assumed that the step size $\{\alpha_{t}\in\text{(0,1]}\}$
is a decreasing sequence satisfying $\alpha_{t}\rightarrow0$, $\sum_{t=1}^{\infty}\alpha_{t}\rightarrow\infty$
and $\sum_{t=1}^{\infty}\alpha_{t}^{2}\rightarrow\infty$. Then, we
have the following Theorem.\begin{theorem}\label{theorem-1} \cite{BSDGD}
	Algorithm \ref{Algorithm-1} can be guaranteed to converge to a stationary
	point when the step size fulfills the condition $\alpha_{t}\in(0,1/L]$,
	where $L$ is the Lipschitz constant. \end{theorem}

\textbf{\textit{Proof: }}Please refer to \cite{BSDGD}.\hspace{12cm}$\blacksquare$

\begin{lemma}\label{lemma-1} The Lipschitz constant of $u_{k}(\mathbf{X},\mathbf{H}_{k}^{(t)})$
is given by 
\begin{align}
L=\begin{cases}
0 & \textrm{if }1-\frac{\Omega_{k}\left(\mathbf{X},\mathbf{H}_{k}^{(t)}\right)}{\omega_{k}}<0\\
\max(l_{e1},l_{A1},l_{D1}) & \textrm{otherwise}\\
\max(l_{e2},l_{A2},l_{D2}) & \textrm{if }1-\frac{\Omega_{k}\left(\mathbf{X},\mathbf{H}_{k}^{(t)}\right)}{\omega_{k}}>\epsilon,
\end{cases}\label{eq:grad-x-2}
\end{align}
where 
\begin{align*}
a & =(UM+1)P_{max}^{2}\lambda_{\max}\left(\mathbf{H}_{k}^{(0),\mathrm{H}}\mathbf{H}_{k}^{(0)}\right)^{2},\\
b & =(UM+1)P_{max}\lambda_{\max}(\mathbf{H}_{k}^{(0),\mathrm{H}}\mathbf{H}_{k}^{(0)}),\\
l_{e1} & =\frac{1}{\omega_{k}^{2}\epsilon\sigma_{k}^{4}}\left((2+5\omega_{k})a-\frac{4ab}{\sigma_{k}^{2}}+\frac{6ab^{2}}{\sigma_{k}^{4}}\right),\\
l_{A1} & =\frac{b^{2}}{\omega_{k}^{2}\epsilon\sigma_{k}^{4}NN_{RF}}\left(2+\frac{(5\omega_{k}-4b)}{\sigma_{k}^{2}}+\frac{6b^{2}}{\sigma_{k}^{4}}\right),\\
l_{D2} & =\frac{NN_{RF}b^{2}}{\omega_{k}^{2}\epsilon\sigma_{k}^{4}P_{max}}\left(2+\omega_{k}+\frac{6b^{2}}{\sigma_{k}^{4}}\right),\\
l_{e2} & =\frac{5a}{\omega_{k}\sigma_{k}^{4}},\\
l_{A2} & =\frac{5b^{2}}{\omega_{k}\sigma_{k}^{4}NN_{RF}},\\
l_{D2} & =\frac{b^{2}NN_{RF}}{\omega_{k}\sigma_{k}^{4}P_{max}}.
\end{align*}
\end{lemma}

\textbf{\textit{Proof: }}See Appendix \ref{subsec:The-proof}.\hspace{12.5cm}$\blacksquare$

\section{Numerical results and discussion}

In this section, numerical results are illustrated to evaluate the
performance of the proposed algorithm. All results are obtained by
averaging over 500 channel realizations. Unless stated otherwise,
we assume $L_{BU}=L_{IU}=L_{BI}=5$, the BS, RIS 1 and RIS 2 are located
at (0 m, 0 m), (40 m, 10 m) and (40 m, -10 m), respectively, and the
users are assumed to be randomly distributed in a circle centered
at (50 m, 0 m) with radius 5 m. The carrier frequency of the mmWave
system is 28 GHz, and the corresponding large-scale fading parameters
are set according to Table I in \cite{mmWave-channel}. Other simulation
parameters are $P_{max}=5$ W, $\sigma_{1}^{2}=,\ldots,\sigma_{K}^{2}=-100$
dBm. For simplicity, it is assumed that the blockage probabilities
are equal $p_{k,l}=p_{\mathrm{block}},\forall k,l$, and that the
target SINR of all the users is $\omega_{1}=\ldots=\omega_{K}=\omega$,
leading to the minimum target rate $R_{\mathrm{targ}}=\log_{2}(1+\omega)$.
To evaluate the performance of the proposed BSGD algorithm, we consider
three benchmark schemes: 1) RIS-random: $\mathbf{e}$ is designed
randomly; 2) RIS-non-robust: the beamforming is designed by setting
the blockage probability to zero; 3) Non-RIS: RISs are not deployed
in the network.

In order to demonstrate the robustness of the proposed algorithm,
we consider two performance metrics: the outage probability and the
effective sum rate. In particular, the outage probability is the average
achievable outage probability for each user, i.e., $\frac{1}{K}\sum_{k\in\mathcal{K}}\mathrm{Pr}\{\Omega_{k}\left(\mathbf{D},\mathbf{A},\mathbf{e}\right)\leq\omega_{k}\}$.
Then, the corresponding effective rate of the $k$-th user is defined
as $R_{\mathrm{eff},k}\triangleq\mathbb{E}[\log_{2}(1+\Omega_{k}(\mathbf{D},\mathbf{A},\mathbf{e}))]$
if $\Omega_{k}(\mathbf{D},\mathbf{A},\mathbf{e})\geq\omega_{k}$ and
$R_{\mathrm{eff},k}\triangleq0$ otherwise.

\begin{figure}
	\centering \subfigure[Outage probability]{ %
		\begin{minipage}[t]{0.495\linewidth}%
			\centering \includegraphics[width=3.5in]{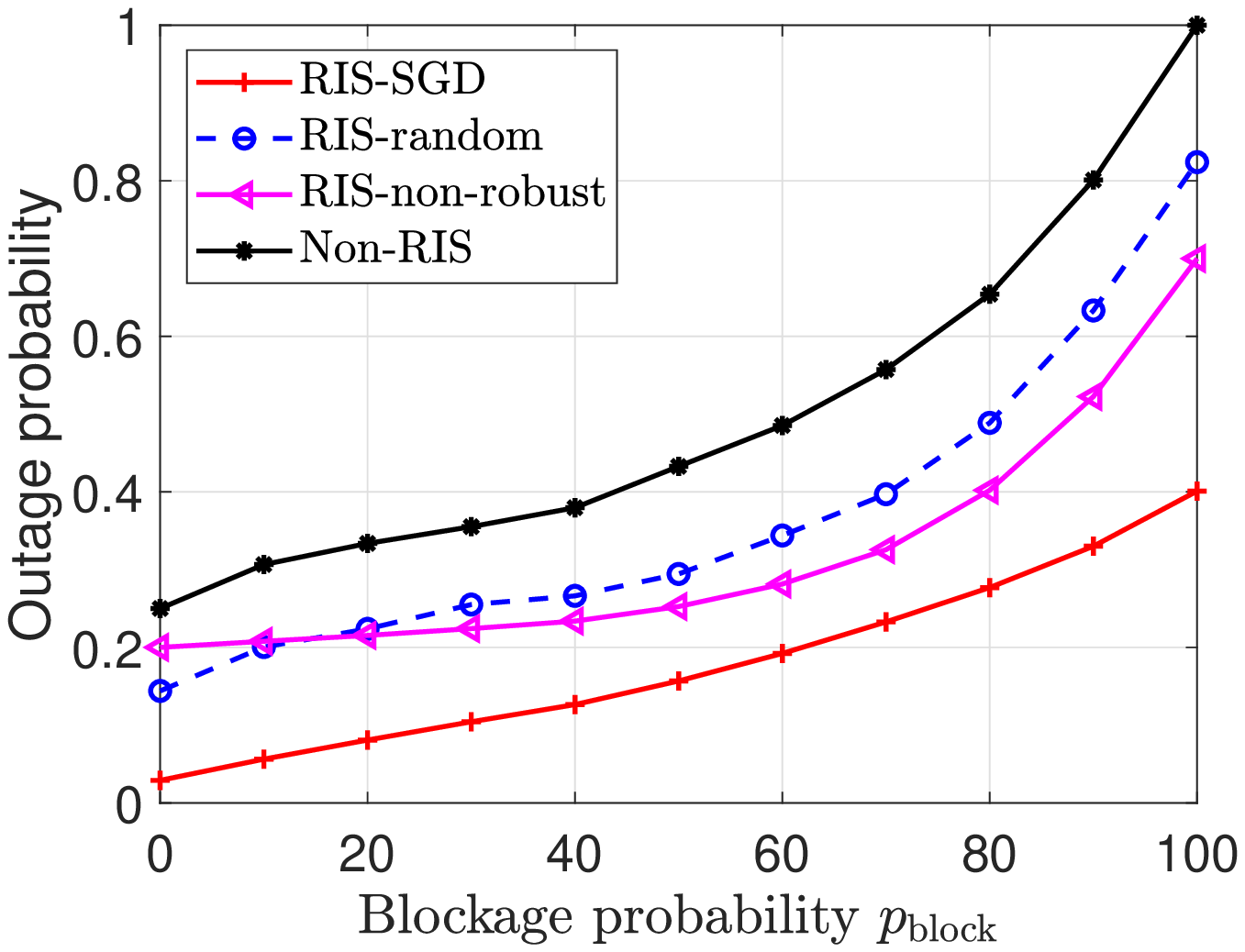} 
	\end{minipage}}\subfigure[Effective sum rate]{ %
		\begin{minipage}[t]{0.495\linewidth}%
			\centering \includegraphics[width=3.5in]{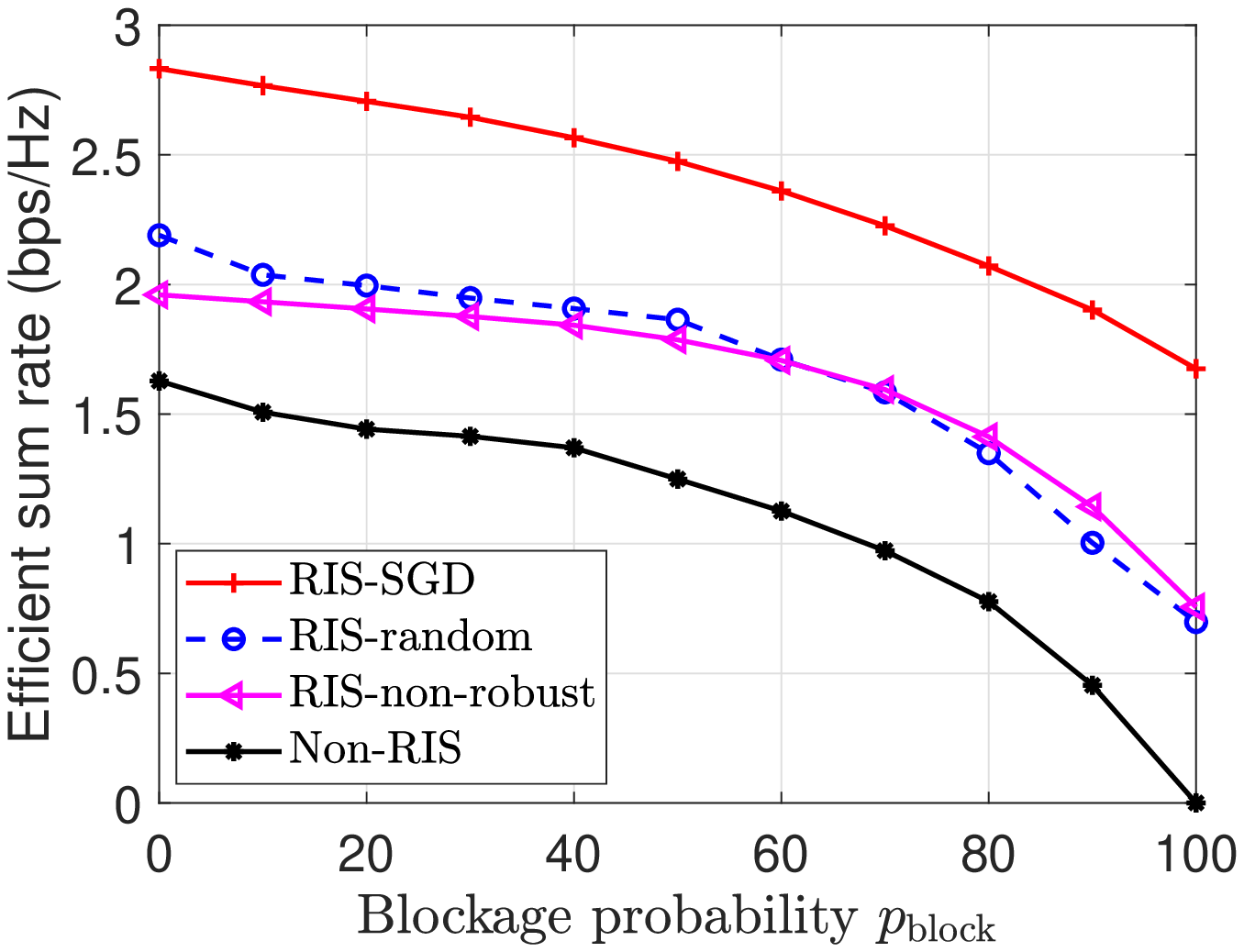} 
	\end{minipage}}
	
	\caption{Outage probability and effective rate as a function of the blockage
		probability $p_{\mathrm{block}}$ for $N=32$, $M=64$, $K=N_{RF}=U=2$,
		and $R_{\mathrm{targ}}=1$ bps/Hz.}
	\label{blockage} 
\end{figure}

Figure \ref{blockage} illustrates the impact of the blockage probability
$p_{\mathrm{block}}$ on the system performance. It is observed that
an mmWave system in the absence of RISs is highly sensitive to the
presence of blockages, which results in the worst outage probability
and effective sum rate. On the other hand, the proposed BSGD-based
scheme provides a high data rate and a low outage probability over
the whole range of blockage probabilities, which substantiates the
robustness of the proposed algorithm to the presence of random blockages.

\begin{figure}
	\centering \includegraphics[width=4in,height=2in]{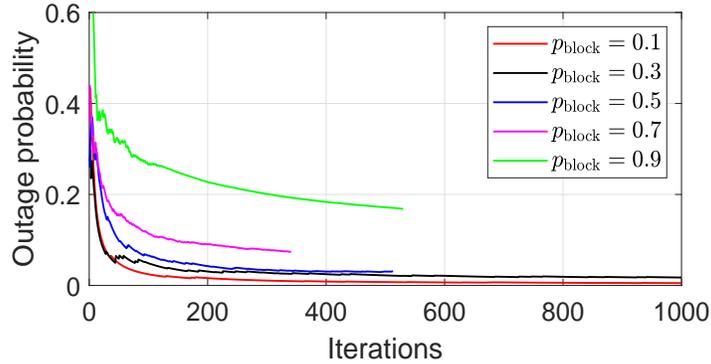}
	\caption{Convergence behavior for $N=32$, $M=64$, $K=N_{RF}=U=2$, and $R_{\mathrm{targ}}=1$
		bps/Hz.}
	\label{converge} 
\end{figure}

Figure \ref{converge} illustrates the convergence behavior of the
stochastic optimization method. It is observed that the BSGD algorithm
converges approximately monotonically in the low outage probability
regime (e.g., $p_{\mathrm{block}}=0.1$), but it shows an oscillatory
behavior in the high outage probability regime (e.g., $p_{\mathrm{block}}=0.9$).
This is due to the nature of stochastic programming.

\section{Conclusions}

This work introduced a stochastic-learning-based robust beamforming
design for RISs-aided mmWave systems which is aimed to combat the
channel uncertainties caused by the presence of random blockages.
The formulated stochastic optimization problem was solved by using
the stochastic block gradient descent algorithm. Simulation results
demonstrated the robustness of the proposed hybrid beamforming design
in the presence of random blockages and confirmed its superior performance
in terms of outage probability and effective data rate compared with
baseline schemes.

\appendices{}

\section{The proof of Lemma \ref{lemma-1}\label{subsec:The-proof}}

A valid choice for Lipschitz constant $L$ is $L\geq\max_{\mathbf{x}\in\mathcal{S}_{x}}||\nabla_{\mathbf{x}}^{2}u_{k}(\mathbf{X},\mathbf{H}_{k})||_{2}$.
However, computing $\max_{\mathbf{x}\in\mathcal{S}_{x}}||\nabla_{\mathbf{x}}^{2}u_{k}(\mathbf{X},\mathbf{H}_{k})||_{2}$
for unstructured $u_{k}(\mathbf{X},\mathbf{H}_{k})$ is not in general
easy. In all such cases, a natural option is leveraging some upper
bound of $\max_{\mathbf{x}\in\mathcal{S}_{x}}||\nabla_{\mathbf{x}}^{2}u_{k}(\mathbf{X},\mathbf{H}_{k})||_{2}$,
that is $\lambda_{\max}(\nabla_{\mathbf{x}}^{2}u_{k}(\mathbf{X},\mathbf{H}_{k}))$.
In the following, we explore to derive the Hessian matrix of $u_{k}(\mathbf{X},\mathbf{H}_{k})$,
and then uniformly bound the largest eigenvalue of the Hessian matrix.

First, the partial derivative and conjugate partial derivative of
$\Omega_{k}\left(\mathbf{X},\mathbf{H}_{k}\right)$ are given by 
\begin{align*}
\frac{\partial\Omega_{k}}{\partial_{\mathbf{x}}} & =\frac{\mathbf{Q}_{\mathbf{x},k}^{T}\mathbf{x}^{*}}{v_{k}}-\frac{\mathbf{x}^{H}\mathbf{Q}_{\mathbf{x},k}\mathbf{x}}{v_{k}^{2}}\overline{\mathbf{Q}}_{\mathbf{x},k}^{T}\mathbf{x}^{*},\\
\frac{\partial\Omega_{k}}{\partial_{\mathbf{x}^{*}}} & =\frac{\mathbf{Q}_{\mathbf{x},k}\mathbf{x}}{v_{k}}-\frac{\mathbf{x}^{H}\mathbf{Q}_{\mathbf{x},k}\mathbf{x}}{v_{k}^{2}}\overline{\mathbf{Q}}_{\mathbf{x},k}\mathbf{x}.
\end{align*}

Second, the partial Hessian matrices of $\Omega_{k}\left(\mathbf{X},\mathbf{H}_{k}\right)$
are given by 
\begin{align*}
\frac{\partial^{2}\Omega_{k}}{\partial_{\mathbf{x}^{*}}\partial_{\mathbf{x}^{T}}} & =\frac{\mathbf{Q}_{\mathbf{x},k}}{v_{k}}-\frac{\overline{\mathbf{Q}}_{\mathbf{x},k}\mathbf{x}\mathbf{x}^{H}\mathbf{Q}_{\mathbf{x},k}}{v_{k}^{2}}-\frac{\mathbf{Q}_{\mathbf{x},k}\mathbf{x}\mathbf{x}^{H}\overline{\mathbf{Q}}_{\mathbf{x},k}}{v_{k}^{2}}-\frac{\mathbf{x}^{H}\mathbf{Q}_{\mathbf{x},k}\mathbf{x}}{v_{k}^{2}}\overline{\mathbf{Q}}_{\mathbf{x},k}+\frac{2\overline{\mathbf{Q}}_{\mathbf{x},k}\mathbf{x}\mathbf{x}^{H}\mathbf{Q}_{\mathbf{x},k}\mathbf{x}\mathbf{x}^{H}\overline{\mathbf{Q}}_{\mathbf{x},k}}{v^{3}},\\
\frac{\partial^{2}\Omega_{k}}{\partial_{\mathbf{x}}\partial_{\mathbf{x}^{H}}} & =\frac{\mathbf{Q}_{\mathbf{x},k}^{T}}{v_{k}}-\frac{\overline{\mathbf{Q}}_{\mathbf{x},k}^{T}\mathbf{x}^{*}\mathbf{x}^{T}\mathbf{Q}_{\mathbf{x},k}^{T}}{v_{k}^{2}}-\frac{\mathbf{Q}_{\mathbf{x},k}^{T}\mathbf{x}^{*}\mathbf{x}^{T}\overline{\mathbf{Q}}_{\mathbf{x},k}^{T}}{v_{k}^{2}}-\frac{\mathbf{x}^{\mathrm{H}}\mathbf{Q}_{\mathbf{x},k}\mathbf{x}}{v_{k}^{2}}\overline{\mathbf{Q}}_{\mathbf{x},k}^{T}\\
 & \ \ \ +\frac{2\overline{\mathbf{Q}}_{\mathbf{x},k}^{T}\mathbf{x}^{*}\mathbf{x}^{H}\mathbf{Q}_{\mathbf{x},k}\mathbf{x}\mathbf{x}^{T}\overline{\mathbf{Q}}_{\mathbf{x},k}^{T}}{v_{k}^{3}},\\
\frac{\partial^{2}\Omega_{k}}{\partial_{\mathbf{x}^{*}}\partial_{\mathbf{x}^{H}}} & =-\frac{\overline{\mathbf{Q}}_{\mathbf{x},k}\mathbf{x}\mathbf{x}^{T}\mathbf{Q}_{\mathbf{x},k}^{T}}{v_{k}^{2}}-\frac{\mathbf{Q}_{\mathbf{x},k}\mathbf{x}\mathbf{x}^{T}\overline{\mathbf{Q}}_{\mathbf{x},k}^{T}}{v_{k}^{2}}+\frac{2\overline{\mathbf{Q}}_{\mathbf{x},k}\mathbf{x}\mathbf{x}^{H}\mathbf{Q}_{\mathbf{x},k}\mathbf{x}\mathbf{x}^{T}\overline{\mathbf{Q}}_{\mathbf{x},k}^{T}}{v_{k}^{3}},\\
\frac{\partial^{2}\Omega_{k}}{\partial_{\mathbf{x}}\partial_{\mathbf{x}^{T}}} & =-\frac{\overline{\mathbf{Q}}_{\mathbf{x},k}^{T}\mathbf{x}^{*}\mathbf{x}^{H}\mathbf{Q}_{\mathbf{x},k}}{v_{k}^{2}}-\frac{\mathbf{Q}_{\mathbf{x},k}^{T}\mathbf{x}^{*}\mathbf{x}^{H}\overline{\mathbf{Q}}_{\mathbf{x},k}}{v_{k}^{2}}+\frac{2\overline{\mathbf{Q}}_{\mathbf{x},k}^{T}\mathbf{x}^{*}\mathbf{x}^{H}\mathbf{Q}_{\mathbf{x},k}\mathbf{x}\mathbf{x}^{H}\overline{\mathbf{Q}}_{\mathbf{x},k}}{v_{k}^{3}}.
\end{align*}

Then, the complex Hessian matrix of $\Omega_{k}\left(\mathbf{X},\mathbf{H}_{k}\right)$
is defined by 
\begin{align*}
\mathcal{H}_{\Omega_{k}} & =\left[\begin{array}{cc}
\frac{\partial^{2}\Omega_{k}}{\partial_{\mathbf{x}^{*}}\partial_{\mathbf{x}^{T}}} & \frac{\partial^{2}\Omega_{k}}{\partial_{\mathbf{x}^{*}}\partial_{\mathbf{x}^{H}}}\\
\frac{\partial^{2}\Omega_{k}}{\partial_{\mathbf{x}}\partial_{\mathbf{x}^{T}}} & \frac{\partial^{2}\Omega_{k}}{\partial_{\mathbf{x}}\partial_{\mathbf{x}^{H}}}
\end{array}\right]\\
 & =\frac{1}{v_{k}}\left[\begin{array}{cc}
\mathbf{Q}_{\mathbf{x},k} & \mathbf{0}\\
\mathbf{0} & \mathbf{Q}_{\mathbf{x},k}^{T}
\end{array}\right]-\frac{2}{v_{k}^{2}}\mathrm{Re}\left(\left[\begin{array}{c}
\overline{\mathbf{Q}}_{\mathbf{x},k}\mathbf{x}\\
\overline{\mathbf{Q}}_{\mathbf{x},k}^{T}\mathbf{x}^{*}
\end{array}\right]\left[\begin{array}{cc}
\mathbf{x}^{H}\mathbf{Q}_{\mathbf{x},k} & \mathbf{x}^{T}\mathbf{Q}_{\mathbf{x},k}^{T}\end{array}\right]\right)\\
 & \ \ \ -\frac{\mathbf{x}^{H}\mathbf{Q}_{\mathbf{x},k}\mathbf{x}}{v_{k}^{2}}\left[\begin{array}{cc}
\overline{\mathbf{Q}}_{\mathbf{x},k} & \mathbf{0}\\
\mathbf{0} & \overline{\mathbf{Q}}_{\mathbf{x},k}^{T}
\end{array}\right]+\frac{2\mathbf{x}^{H}\mathbf{Q}_{\mathbf{x},k}\mathbf{x}}{v_{k}^{3}}\left[\begin{array}{c}
\overline{\mathbf{Q}}_{\mathbf{x},k}\mathbf{x}\\
\overline{\mathbf{Q}}_{\mathbf{x},k}^{T}\mathbf{x}^{*}
\end{array}\right]\left[\begin{array}{cc}
\mathbf{x}^{H}\overline{\mathbf{Q}}_{\mathbf{x},k} & \mathbf{x}^{T}\overline{\mathbf{Q}}_{\mathbf{x},k}^{T}\end{array}\right].
\end{align*}

$\bullet$ If $1-\frac{\Omega_{k}\left(\mathbf{X},\mathbf{H}_{k}\right)}{\omega_{k}}>\epsilon$,
then $u_{k}\left(\mathbf{X},\mathbf{H}_{k}\right)=1-\frac{\Omega_{k}\left(\mathbf{X},\mathbf{H}_{k}\right)}{\omega_{k}}-\frac{\epsilon}{2}$,
the complex Hessian matrix of wich is $\mathcal{H}_{u_{k}}=-\frac{\mathcal{H}_{\Omega_{k}}}{\omega_{k}}$.
The bound of $\lambda_{\max}(\mathcal{H}_{u_{k}})$ is given by 
\begin{align}
\lambda_{\max}(\mathcal{H}_{u_{k}}) & \leq\lambda_{\max}\left(-\frac{1}{\omega_{k}v_{k}}\left[\begin{array}{cc}
\mathbf{Q}_{\mathbf{x},k} & \mathbf{0}\\
\mathbf{0} & \mathbf{Q}_{\mathbf{x},k}^{T}
\end{array}\right]\right)+\lambda_{\max}\left(\frac{1}{\omega_{k}v_{k}^{2}}\left[\begin{array}{c}
\overline{\mathbf{Q}}_{\mathbf{x},k}\mathbf{x}\\
\overline{\mathbf{Q}}_{\mathbf{x},k}^{T}\mathbf{x}^{*}
\end{array}\right]\left[\begin{array}{cc}
\mathbf{x}^{H}\mathbf{Q}_{\mathbf{x},k} & \mathbf{x}^{T}\mathbf{Q}_{\mathbf{x},k}^{T}\end{array}\right]\right)\nonumber \\
 & \ \ \ +\lambda_{\max}\left(\frac{1}{\omega_{k}v_{k}^{2}}\left[\begin{array}{c}
\mathbf{Q}_{\mathbf{x},k}\mathbf{x}\\
\mathbf{Q}_{\mathbf{x},k}^{T}\mathbf{x}^{*}
\end{array}\right]\left[\begin{array}{cc}
\mathbf{x}^{\mathrm{H}}\overline{\mathbf{Q}}_{\mathbf{x},k} & \mathbf{x}^{T}\overline{\mathbf{Q}}_{\mathbf{x},k}^{T}\end{array}\right]\right)\nonumber \\
 & \ \ \ +\lambda_{\max}\left(\frac{\mathbf{x}^{\mathrm{H}}\mathbf{Q}_{\mathbf{x},k}\mathbf{x}}{\omega_{k}v_{k}^{2}}\left[\begin{array}{cc}
\overline{\mathbf{Q}}_{\mathbf{x},k} & \mathbf{0}\\
\mathbf{0} & \overline{\mathbf{Q}}_{\mathbf{x},k}^{T}
\end{array}\right]\right)\nonumber \\
 & =\lambda_{\max}\left(-\frac{1}{\omega_{k}v_{k}}\left[\begin{array}{cc}
\mathbf{Q}_{\mathbf{x},k} & \mathbf{0}\\
\mathbf{0} & \mathbf{Q}_{\mathbf{x},k}^{T}
\end{array}\right]\right)+\lambda_{\max}\left(\frac{2}{\omega_{k}v_{k}^{2}}\mathrm{Re}\left(\left[\begin{array}{c}
\overline{\mathbf{Q}}_{\mathbf{x},k}\mathbf{x}\\
\overline{\mathbf{Q}}_{\mathbf{x},k}^{T}\mathbf{x}^{*}
\end{array}\right]\left[\begin{array}{cc}
\mathbf{x}^{H}\mathbf{Q}_{\mathbf{x},k} & \mathbf{x}^{T}\mathbf{Q}_{\mathbf{x},k}^{T}\end{array}\right]\right)\right)\\
 & \ \ \ +\lambda_{\max}\left(\frac{\mathbf{x}^{\mathrm{H}}\mathbf{Q}_{\mathbf{x},k}\mathbf{x}}{\omega_{k}v_{k}^{2}}\left[\begin{array}{cc}
\overline{\mathbf{Q}}_{\mathbf{x},k} & \mathbf{0}\\
\mathbf{0} & \overline{\mathbf{Q}}_{\mathbf{x},k}^{T}
\end{array}\right]\right)\\
 & =-\frac{1}{\omega_{k}v_{k}}\lambda_{\min}\left(\mathbf{Q}_{\mathbf{x},k}\right)+\frac{1}{\omega_{k}v_{k}^{2}}2\max\left(\mathrm{Re}\left(\mathbf{x}^{\mathrm{H}}\mathbf{Q}_{\mathbf{x},k}\overline{\mathbf{Q}}_{\mathbf{x},k}\mathbf{x}+\mathbf{x}^{T}\mathbf{Q}_{\mathbf{x},k}^{T}\overline{\mathbf{Q}}_{\mathbf{x},k}^{T}\mathbf{x}^{*}\right),0\right)\nonumber \\
 & \ \ \ +\frac{\mathbf{x}^{\mathrm{H}}\mathbf{Q}_{\mathbf{x},k}\mathbf{x}}{\omega_{k}v_{k}^{2}}\lambda_{\max}\left(\overline{\mathbf{Q}}_{\mathbf{x},k}\right)\nonumber \\
 & =\frac{4}{\omega_{k}v_{k}^{2}}\max\left(\mathrm{Re}\left(\mathbf{x}^{\mathrm{H}}\mathbf{Q}_{\mathbf{x},k}\overline{\mathbf{Q}}_{\mathbf{x},k}\mathbf{x}\right),0\right)-\frac{1}{\omega_{k}v_{k}}\lambda_{\min}\left(\mathbf{Q}_{\mathbf{x},k}\right)+\frac{\mathbf{x}^{H}\mathbf{Q}_{\mathbf{x},k}\mathbf{x}}{\omega_{k}v_{k}^{2}}\lambda_{\max}\left(\overline{\mathbf{Q}}_{\mathbf{x},k}\right)\nonumber \\
 & \overset{\mathrm{(A1)}}{=}\frac{4}{\omega_{k}v_{k}^{2}}\max\left(\mathrm{Re}\left(\mathbf{x}^{\mathrm{H}}\mathbf{Q}_{\mathbf{x},k}\overline{\mathbf{Q}}_{\mathbf{x},k}\mathbf{x}\right),0\right)+\frac{\mathbf{x}^{H}\mathbf{Q}_{\mathbf{x},k}\mathbf{x}}{\omega_{k}v_{k}^{2}}\lambda_{\max}(\overline{\mathbf{Q}}_{\mathbf{x},k})\nonumber \\
 & \overset{\mathrm{(A2)}}{\leq}\begin{cases}
\frac{5a}{\omega_{k}\sigma_{k}^{4}} & \textrm{if }\mathbf{x}=\mathbf{e}\\
\frac{5b^{2}}{\omega_{k}\sigma_{k}^{4}NN_{RF}} & \textrm{\textrm{if }}\mathbf{x}=\mathbf{\mathrm{vec}({\bf A})}\\
\frac{b^{2}NN_{RF}}{\omega_{k}\sigma_{k}^{4}P_{max}} & \textrm{if }\mathbf{x}=\mathrm{vec}({\bf D}).
\end{cases}\label{eq:LAMDA-1}
\end{align}
where $a=(UM+1)P_{max}^{2}h_{k}^{2}$, $b=(UM+1)P_{max}h_{k}$, and
$h_{k}=\lambda_{\max}(\mathbf{H_{\mathrm{bi}}}^{\mathrm{H}}\mathrm{diag}(\mathbf{h}_{\mathrm{i},k})\mathrm{diag}(\mathbf{h}_{\mathrm{i},k}^{\mathrm{H}})\mathbf{H_{\mathrm{bi}}})+\frac{1}{L_{BU}}\lambda_{\max}(\mathbf{G}_{k}^{H}\mathbf{G}_{k})||\mathbf{g}_{k}||^{2}$.

Equation (A1) in (\ref{eq:LAMDA-1}) is due to the fact that $\mathbf{Q}_{\mathbf{x},k}$
is a rank-1 positive semidefinite metrix. The inequality (A2) in (\ref{eq:LAMDA-1})
comes from $v_{k}=\mathbf{x}^{\mathrm{H}}\overline{\mathbf{Q}}_{\mathbf{x},k}\mathbf{x}+\sigma_{k}^{2}\geq\sigma_{k}^{2}$
and the following relaxations.

\textbf{1) For $\mathrm{Re}\left(\mathbf{x}^{\mathrm{H}}\mathbf{Q}_{\mathbf{x},k}\overline{\mathbf{Q}}_{\mathbf{x},k}\mathbf{x}\right)$:}

\begin{align*}
 & \mathrm{Re}\left(\mathbf{e}^{\mathrm{H}}\mathbf{Q}_{\mathbf{e},k}\overline{\mathbf{Q}}_{\mathbf{e},k}\mathbf{e}\right)\\
 & =\mathrm{Re}\left(\mathrm{Tr}\left(\mathbf{H}_{k}\mathbf{A}{\bf d}_{k}{\bf d}_{k}^{\mathrm{H}}\mathbf{A}^{\mathrm{H}}\mathbf{H}_{k}^{\mathrm{H}}\mathbf{H}_{k}\mathbf{A}\sum_{i\neq k}^{K}{\bf d}_{i}{\bf d}_{i}^{\mathrm{H}}\mathbf{A}^{\mathrm{H}}\mathbf{H}_{k}^{\mathrm{H}}\mathbf{e}\mathbf{e}^{\mathrm{H}}\right)\right)\\
 & \leq(UM+1)\mathrm{Re}\left(\mathrm{Tr}\left(\mathbf{A}{\bf d}_{k}{\bf d}_{k}^{\mathrm{H}}\mathbf{A}^{\mathrm{H}}\mathbf{H}_{k}^{\mathrm{H}}\mathbf{H}_{k}\mathbf{A}\sum_{i\neq k}^{K}{\bf d}_{i}{\bf d}_{i}^{\mathrm{H}}\mathbf{A}^{\mathrm{H}}\mathbf{H}_{k}^{\mathrm{H}}\mathbf{H}_{k}\right)\right)\\
 & \leq(UM+1)\lambda_{\max}\left(\mathbf{H}_{k}^{\mathrm{H}}\mathbf{H}_{k}\right)^{2}\mathrm{Tr}\left(\mathbf{A}{\bf d}_{k}{\bf d}_{k}^{\mathrm{H}}\mathbf{A}^{\mathrm{H}}\mathbf{A}\sum_{i\neq k}^{K}{\bf d}_{i}{\bf d}_{i}^{\mathrm{H}}\mathbf{A}^{\mathrm{H}}\right)\\
 & \leq(UM+1)P_{max}^{2}\lambda_{\max}\left(\mathbf{H}_{k}^{\mathrm{H}}\mathbf{H}_{k}\right)^{2}\\
 & \overset{\mathrm{(A3)}}{\leq}(UM+1)P_{max}^{2}h_{k}^{2}\\
 & =a,
\end{align*}
\begin{align*}
 & \mathrm{Re}\left(\mathrm{vec}({\bf A})^{\mathrm{H}}\mathbf{Q}_{\mathbf{A},k}\overline{\mathbf{Q}}_{\mathbf{A},k}\mathrm{vec}({\bf A})\right)\\
 & =\mathrm{Re}\left(\mathrm{vec}({\bf A})^{\mathrm{H}}\left(\mathbf{d}_{k}^{*}\mathbf{d}_{k}^{\mathrm{T}}\otimes\mathbf{H}_{k}^{\mathrm{H}}\mathbf{e}\mathbf{e}^{\mathrm{H}}\mathbf{H}_{k}\right)\left(\mathbf{D}_{-k}^{*}\mathbf{D}_{-k}^{\mathrm{T}}\otimes\mathbf{H}_{k}^{\mathrm{H}}\mathbf{e}\mathbf{e}^{\mathrm{H}}\mathbf{H}_{k}\right)\mathrm{vec}({\bf A})\right)\\
 & =\mathrm{Re}\left(\mathrm{vec}({\bf A})^{\mathrm{H}}\left(\mathbf{d}_{k}^{*}\mathbf{d}_{k}^{\mathrm{T}}\mathbf{D}_{-k}^{*}\mathbf{D}_{-k}^{\mathrm{T}}\otimes\mathbf{H}_{k}^{\mathrm{H}}\mathbf{e}\mathbf{e}^{\mathrm{H}}\mathbf{H}_{k}\mathbf{H}_{k}^{\mathrm{H}}\mathbf{e}\mathbf{e}^{\mathrm{H}}\mathbf{H}_{k}\right)\mathrm{vec}({\bf A})\right)\\
 & =\mathrm{Re}\left(\mathrm{Tr}\left(\mathbf{H}_{k}^{\mathrm{H}}\mathbf{e}\mathbf{e}^{\mathrm{H}}\mathbf{H}_{k}\mathbf{H}_{k}^{\mathrm{H}}\mathbf{e}\mathbf{e}^{\mathrm{H}}\mathbf{H}_{k}{\bf A}\mathbf{D}_{-k}\mathbf{D}_{-k}^{\mathrm{H}}\mathbf{d}_{k}\mathbf{d}_{k}^{\mathrm{H}}{\bf A}^{\mathrm{H}}\right)\right)\\
 & \leq(UM+1)^{2}\mathrm{Re}\left(\mathrm{Tr}\left(\mathbf{H}_{k}^{\mathrm{H}}\mathbf{H}_{k}\mathbf{H}_{k}^{\mathrm{H}}\mathbf{H}_{k}{\bf A}\mathbf{D}_{-k}\mathbf{D}_{-k}^{\mathrm{H}}\mathbf{d}_{k}\mathbf{d}_{k}^{\mathrm{H}}{\bf A}^{\mathrm{H}}\right)\right)\\
 & \leq(UM+1)^{2}\lambda_{\max}\left(\mathbf{H}_{k}^{\mathrm{H}}\mathbf{H}_{k}\right)^{2}\mathrm{Re}\left(\mathrm{Tr}\left({\bf A}\mathbf{D}_{-k}\mathbf{D}_{-k}^{\mathrm{H}}\mathbf{d}_{k}\mathbf{d}_{k}^{\mathrm{H}}{\bf A}^{\mathrm{H}}\right)\right)\\
 & \leq\frac{(UM+1)^{2}P_{max}^{2}}{NN_{RF}}\lambda_{\max}\left(\mathbf{H}_{k}^{\mathrm{H}}\mathbf{H}_{k}\right)^{2}\\
 & \overset{\mathrm{(A4)}}{\leq}\frac{(UM+1)^{2}P_{max}^{2}}{NN_{RF}}h_{k}^{2}\\
 & =\frac{b^{2}}{NN_{RF}},
\end{align*}
and 
\begin{align*}
 & \mathrm{Re}\left(\mathrm{vec}({\bf D})^{\mathrm{H}}\mathbf{Q}_{\mathbf{D},k}\overline{\mathbf{Q}}_{\mathbf{D},k}\mathrm{vec}({\bf D})\right)\\
 & =\mathrm{Re}\left(\mathrm{vec}({\bf D})^{\mathrm{H}}\left(\textrm{diag}(\mathbf{i}_{k})\otimes\mathbf{A}^{\mathrm{H}}\mathbf{H}_{k}^{\mathrm{H}}\mathbf{e}\mathbf{e}^{\mathrm{H}}\mathbf{H}_{k}\mathbf{A}\right)\left(\textrm{diag}(\mathbf{\overline{i}}_{k})\otimes\mathbf{A}^{\mathrm{H}}\mathbf{H}_{k}^{\mathrm{H}}\mathbf{e}\mathbf{e}^{\mathrm{H}}\mathbf{H}_{k}\mathbf{A}\right)\mathrm{vec}({\bf D})\right)\\
 & =\mathrm{Re}\left(\mathrm{vec}({\bf D})^{\mathrm{H}}\left(\textrm{diag}(\mathbf{i}_{k})\textrm{diag}(\mathbf{\overline{i}}_{k})\otimes\mathbf{A}^{\mathrm{H}}\mathbf{H}_{k}^{\mathrm{H}}\mathbf{e}\mathbf{e}^{\mathrm{H}}\mathbf{H}_{k}\mathbf{A}\mathbf{A}^{\mathrm{H}}\mathbf{H}_{k}^{\mathrm{H}}\mathbf{e}\mathbf{e}^{\mathrm{H}}\mathbf{H}_{k}\mathbf{A}\right)\mathrm{vec}({\bf D})\right)\\
 & =0,
\end{align*}
where inequality (A3) and (A4) is because the maximum gain of the
random channel $\mathbf{H}_{k}$ is obtained at the blockage probability
of $p_{k,l}=0,\forall k,l$. Specifically, (\ref{eq:5-1}) can be
rewritten as

\begin{equation}
\mathbf{h}_{\mathrm{b},k}=\sqrt{\frac{1}{L_{BU}}}\mathbf{G}_{k}\boldsymbol{\gamma}_{k},\forall k\in\mathcal{K},\label{eq:5-1-1}
\end{equation}
where $\mathbf{G}_{k}=[\mathbf{a}_{L}(\theta_{k,1}^{\mathrm{b},t}),\ldots,\mathbf{a}_{L}(\theta_{k,L_{BU}}^{\mathrm{b},t})]$
and $\boldsymbol{\gamma}_{k}=[\gamma_{k,1}g_{k,1}^{\mathrm{b}},\ldots,\gamma_{k,L_{BU}}g_{k,L_{BU}}^{\mathrm{b}}]^{T}$.
Then, based on channel model (\ref{eq:5-1-1}), we have the following
inequaity

\begin{align*}
\lambda_{\max}\left(\mathbf{H}_{k}^{\mathrm{H}}\mathbf{H}_{k}\right) & =\underbrace{\lambda_{\max}\left(\mathbf{H_{\mathrm{bi}}}^{\mathrm{H}}\mathrm{diag}(\mathbf{h}_{\mathrm{i},k})\mathrm{diag}(\mathbf{h}_{\mathrm{i},k}^{\mathrm{H}})\mathbf{H_{\mathrm{bi}}}\right)}_{\textrm{deterministic term}}+\underbrace{\lambda_{\max}\left(\mathbf{h}_{\mathrm{b},k}\mathbf{h}_{\mathrm{b},k}^{H}\right)}_{\textrm{stochastic term}}\\
 & \leq\underbrace{\lambda_{\max}\left(\mathbf{H_{\mathrm{bi}}}^{\mathrm{H}}\mathrm{diag}(\mathbf{h}_{\mathrm{i},k})\mathrm{diag}(\mathbf{h}_{\mathrm{i},k}^{\mathrm{H}})\mathbf{H_{\mathrm{bi}}}\right)}_{\textrm{deterministic term}}+\underbrace{\frac{1}{L_{BU}}\lambda_{\max}\left(\mathbf{G}_{k}^{H}\mathbf{G}_{k}\right)||\boldsymbol{\gamma}_{k}||^{2}}_{\textrm{stochastic term}}\\
 & \leq\underbrace{\lambda_{\max}\left(\mathbf{H_{\mathrm{bi}}}^{\mathrm{H}}\mathrm{diag}(\mathbf{h}_{\mathrm{i},k})\mathrm{diag}(\mathbf{h}_{\mathrm{i},k}^{\mathrm{H}})\mathbf{H_{\mathrm{bi}}}\right)}_{\textrm{deterministic term}}+\underbrace{\frac{1}{L_{BU}}\lambda_{\max}\left(\mathbf{G}_{k}^{H}\mathbf{G}_{k}\right)||\mathbf{g}_{k}||^{2}}_{\textrm{deterministic term}}\\
 & =h_{k}
\end{align*}
where $\mathbf{g}_{k}=[g_{k,1}^{\mathrm{b}},\ldots,g_{k,L_{BU}}^{\mathrm{b}}]^{T}$.

\textbf{2) For $\lambda_{\max}(\overline{\mathbf{Q}}_{\mathbf{x},k})$:}

\begin{align*}
\mathbf{x}^{\mathrm{H}}\mathbf{Q}_{\mathbf{x},k}\mathbf{x} & \leq\lambda_{\max}(\mathbf{H}_{k}^{(t),\mathrm{H}}\mathbf{H}_{k}^{(t)})\lambda_{\max}({\bf e}{\bf e}^{\mathrm{H}})\mathrm{Tr}(\mathbf{A}{\bf d}_{k}{\bf d}_{k}^{\mathrm{H}}\mathbf{A}^{\mathrm{H}})\leq b,\\
\mathbf{x}^{\mathrm{H}}\overline{\mathbf{Q}}_{\mathbf{x},k}^{(t)}\mathbf{x} & \leq\lambda_{\max}(\mathbf{H}_{k}^{(t),\mathrm{H}}\mathbf{H}_{k}^{(t)})\lambda_{\max}({\bf e}{\bf e}^{\mathrm{H}})\mathrm{Tr}(\mathbf{A}\sum_{i\neq k}^{K}{\bf d}_{i}{\bf d}_{i}^{\mathrm{H}}\mathbf{A}^{\mathrm{H}})\leq b,\\
\lambda_{\max}(\overline{\mathbf{Q}}_{\mathbf{e},k}) & \leq\frac{b}{\mathrm{Tr}({\bf e}{\bf e}^{\mathrm{H}})}\leq\frac{b}{UM+1},\\
\lambda_{\max}(\overline{\mathbf{Q}}_{\mathbf{A},k}^{(t)}) & \leq\frac{b}{\mathrm{Tr}({\bf A}{\bf A}^{\mathrm{H}})}\leq\frac{b}{NN_{RF}},\\
\lambda_{\max}(\overline{\mathbf{Q}}_{\mathbf{D},k}^{(t)}) & \leq\frac{b}{\mathrm{Tr}({\bf D}{\bf D}^{\mathrm{H}})}\leq\frac{b\mathrm{Tr}({\bf A}{\bf A}^{\mathrm{H}})}{P_{max}}\leq\frac{bNN_{RF}}{P_{max}}.
\end{align*}

$\bullet$ If $0\leq1-\frac{\Omega_{k}\left(\mathbf{X},\mathbf{H}_{k}\right)}{\omega_{k}}\leq\epsilon$,
then $u_{k}\left(\mathbf{X},\mathbf{H}_{k}\right)=\frac{1}{2\epsilon}\left(1-\frac{\Omega_{k}\left(\mathbf{X},\mathbf{H}_{k}\right)}{\omega_{k}}\right)^{2}$.
The partial Hessian matrices of $u_{k}\left(\mathbf{X},\mathbf{H}_{k}\right)$
are given by 
\begin{align*}
\frac{\partial^{2}u_{k}}{\partial_{\mathbf{x}^{*}}\partial_{\mathbf{x}^{T}}} & =\frac{1}{\omega_{k}^{2}\epsilon}\left(\frac{\partial\Omega_{k}}{\partial_{\mathbf{x}^{*}}}\frac{\partial\Omega_{k}}{\partial_{\mathbf{x}^{T}}}+(\Omega_{k}-\omega_{k})\frac{\partial^{2}\Omega_{k}}{\partial_{\mathbf{x}^{*}}\partial_{\mathbf{x}^{T}}}\right),\\
\frac{\partial^{2}u_{k}}{\partial_{\mathbf{x}}\partial_{\mathbf{x}^{H}}} & =\frac{1}{\omega_{k}^{2}\epsilon}\left(\frac{\partial\Omega_{k}}{\partial_{\mathbf{x}}}\frac{\partial\Omega_{k}}{\partial_{\mathbf{x}^{H}}}+(\Omega_{k}-\omega_{k})\frac{\partial^{2}\Omega_{k}}{\partial_{\mathbf{x}}\partial_{\mathbf{x}^{H}}}\right),\\
\frac{\partial^{2}\Omega_{k}}{\partial_{\mathbf{x}^{*}}\partial_{\mathbf{x}^{H}}} & =\frac{1}{\omega_{k}^{2}\epsilon}\left(\frac{\partial\Omega_{k}}{\partial_{\mathbf{x}^{*}}}\frac{\partial\Omega_{k}}{\partial_{\mathbf{x}^{H}}}+(\Omega_{k}-\omega_{k})\frac{\partial^{2}\Omega_{k}}{\partial_{\mathbf{x}^{*}}\partial_{\mathbf{x}^{H}}}\right),\\
\frac{\partial^{2}\Omega_{k}}{\partial_{\mathbf{x}}\partial_{\mathbf{x}^{T}}} & =\frac{1}{\omega_{k}^{2}\epsilon}\left(\frac{\partial\Omega_{k}}{\partial_{\mathbf{x}}}\frac{\partial\Omega_{k}}{\partial_{\mathbf{x}^{T}}}+(\Omega_{k}-\omega_{k})\frac{\partial^{2}\Omega_{k}}{\partial_{\mathbf{x}}\partial_{\mathbf{x}^{T}}}\right).
\end{align*}

The bound of $\lambda_{\max}(\mathcal{H}_{u_{k}})$ is given by

\begin{align}
\lambda_{\max}(\mathcal{H}_{u_{k}}) & =\lambda_{\max}\left(\left[\begin{array}{cc}
\frac{\partial^{2}u_{k}}{\partial_{\mathbf{x}^{*}}\partial_{\mathbf{x}^{T}}} & \frac{\partial^{2}u_{k}}{\partial_{\mathbf{x}^{*}}\partial_{\mathbf{x}^{H}}}\\
\frac{\partial^{2}u_{k}}{\partial_{\mathbf{x}}\partial_{\mathbf{x}^{T}}} & \frac{\partial^{2}u_{k}}{\partial_{\mathbf{x}}\partial_{\mathbf{x}^{H}}}
\end{array}\right]\right)\nonumber \\
 & \leq\frac{1}{\omega_{k}^{2}\epsilon}\lambda_{\max}\left(\left[\begin{array}{cc}
\frac{\partial\Omega_{k}}{\partial_{\mathbf{x}^{*}}}\frac{\partial\Omega_{k}}{\partial_{\mathbf{x}^{T}}} & \frac{\partial\Omega_{k}}{\partial_{\mathbf{x}^{*}}}\frac{\partial\Omega_{k}}{\partial_{\mathbf{x}^{H}}}\\
\frac{\partial\Omega_{k}}{\partial_{\mathbf{x}}}\frac{\partial\Omega_{k}}{\partial_{\mathbf{x}^{T}}} & \frac{\partial\Omega_{k}}{\partial_{\mathbf{x}}}\frac{\partial\Omega_{k}}{\partial_{\mathbf{x}^{H}}}
\end{array}\right]\right)+\frac{1}{\omega_{k}^{2}\epsilon}\lambda_{\max}\left((\Omega_{k}-\omega_{k})\mathcal{H}_{\Omega_{k}}\right)\nonumber \\
 & \leq\frac{1}{\omega_{k}^{2}\epsilon}\lambda_{\max}\left(\left[\begin{array}{c}
\frac{\partial\Omega_{k}}{\partial_{\mathbf{x}^{*}}}\\
\frac{\partial\Omega_{k}}{\partial_{\mathbf{x}}}
\end{array}\right]\left[\begin{array}{cc}
\frac{\partial\Omega_{k}}{\partial_{\mathbf{x}^{T}}} & \frac{\partial\Omega_{k}}{\partial_{\mathbf{x}^{H}}}\end{array}\right]\right)+\frac{1}{\omega_{k}^{2}\epsilon}\Omega_{k}\lambda_{\max}\left(\mathcal{H}_{\Omega_{k}}\right)+\frac{1}{\epsilon}\lambda_{\max}(\mathcal{H}_{u_{k}}),\label{eq:H-1}
\end{align}
where

\begin{align}
 & \lambda_{\max}\left(\left[\begin{array}{c}
\frac{\partial\Omega_{k}}{\partial_{\mathbf{x}^{*}}}\\
\frac{\partial\Omega_{k}}{\partial_{\mathbf{x}}}
\end{array}\right]\left[\begin{array}{cc}
\frac{\partial\Omega_{k}}{\partial_{\mathbf{x}^{T}}} & \frac{\partial\Omega_{k}}{\partial_{\mathbf{x}^{H}}}\end{array}\right]\right)\nonumber \\
= & 2\mathrm{Re}\left(\frac{\partial\Omega_{k}}{\partial_{\mathbf{x}^{T}}}\frac{\partial\Omega_{k}}{\partial_{\mathbf{x}^{*}}}\right)\\
= & 2\mathrm{Re}\left(\left(\frac{\mathbf{x}^{H}\mathbf{Q}_{\mathbf{x},k}}{v_{k}}-\frac{\mathbf{x}^{H}\mathbf{Q}_{\mathbf{x},k}\mathbf{x}}{v_{k}^{2}}\mathbf{x}^{H}\mathbf{\overline{Q}}_{\mathbf{x},k}\right)\left(\frac{\mathbf{Q}_{\mathbf{x},k}\mathbf{x}}{v_{k}}-\frac{\mathbf{x}^{H}\mathbf{Q}_{\mathbf{x},k}\mathbf{x}}{v_{k}^{2}}\overline{\mathbf{Q}}_{\mathbf{x},k}\mathbf{x}\right)\right)\nonumber \\
= & \frac{\mathbf{x}^{H}\mathbf{Q}_{\mathbf{x},k}\mathbf{Q}_{\mathbf{x},k}\mathbf{x}}{v_{k}^{2}}-4\mathrm{Re}\left(\frac{\mathbf{x}^{H}\mathbf{Q}_{\mathbf{x},k}\mathbf{x}\mathbf{x}^{H}\mathbf{Q}_{\mathbf{x},k}\overline{\mathbf{Q}}_{\mathbf{x},k}\mathbf{x}}{v_{k}^{3}}\right)+\frac{\left(\mathbf{x}^{H}\mathbf{Q}_{\mathbf{x},k}\mathbf{x}\right)^{2}}{v_{k}^{4}}\mathbf{x}^{H}\mathbf{\overline{Q}}_{\mathbf{x},k}\overline{\mathbf{Q}}_{\mathbf{x},k}\mathbf{x},\label{eq:H-2}
\end{align}
and

\begin{align}
 & \lambda_{\max}\left(\mathcal{H}_{\Omega_{k}}\right)\nonumber \\
\leq & \lambda_{\max}\left(\frac{1}{v_{k}}\left[\begin{array}{cc}
\mathbf{Q}_{\mathbf{x},k} & \mathbf{0}\\
\mathbf{0} & \mathbf{Q}_{\mathbf{x},k}^{T}
\end{array}\right]\right)+\lambda_{\max}\left(-\frac{2}{v_{k}^{2}}\mathrm{Re}\left(\left[\begin{array}{c}
\overline{\mathbf{Q}}_{\mathbf{x},k}\mathbf{x}\\
\overline{\mathbf{Q}}_{\mathbf{x},k}^{T}\mathbf{x}^{*}
\end{array}\right]\left[\begin{array}{cc}
\mathbf{x}^{H}\mathbf{Q}_{\mathbf{x},k} & \mathbf{x}^{T}\mathbf{Q}_{\mathbf{x},k}^{T}\end{array}\right]\right)\right)\\
 & +\lambda_{\max}\left(-\frac{\mathbf{x}^{\mathrm{H}}\mathbf{Q}_{\mathbf{x},k}\mathbf{x}}{\omega_{k}v_{k}^{2}}\left[\begin{array}{cc}
\overline{\mathbf{Q}}_{\mathbf{x},k} & \mathbf{0}\\
\mathbf{0} & \overline{\mathbf{Q}}_{\mathbf{x},k}^{T}
\end{array}\right]\right)\nonumber \\
 & +\lambda_{\max}\left(\frac{2\mathbf{x}^{H}\mathbf{Q}_{\mathbf{x},k}\mathbf{x}}{v_{k}^{3}}\left[\begin{array}{c}
\overline{\mathbf{Q}}_{\mathbf{x},k}\mathbf{x}\\
\overline{\mathbf{Q}}_{\mathbf{x},k}^{T}\mathbf{x}^{*}
\end{array}\right]\left[\begin{array}{cc}
\mathbf{x}^{H}\overline{\mathbf{Q}}_{\mathbf{x},k} & \mathbf{x}^{T}\overline{\mathbf{Q}}_{\mathbf{x},k}^{T}\end{array}\right]\right)\nonumber \\
= & \frac{1}{v_{k}}\lambda_{\max}\left(\mathbf{Q}_{\mathbf{x},k}\right)+\frac{4}{v_{k}^{2}}\max\left(-\mathrm{Re}\left(\mathbf{x}^{\mathrm{H}}\mathbf{Q}_{\mathbf{x},k}\overline{\mathbf{Q}}_{\mathbf{x},k}\mathbf{x}\right),0\right)\nonumber \\
 & -\frac{\mathbf{x}^{\mathrm{H}}\mathbf{Q}_{\mathbf{x},k}\mathbf{x}}{\omega_{k}v_{k}^{2}}\lambda_{\min}\left(\overline{\mathbf{Q}}_{\mathbf{x},k}\right)+\frac{4\mathbf{x}^{H}\mathbf{Q}_{\mathbf{x},k}\mathbf{x}}{v_{k}^{3}}\mathbf{x}^{H}\overline{\mathbf{Q}}_{\mathbf{x},k}\overline{\mathbf{Q}}_{\mathbf{x},k}\mathbf{x}\nonumber \\
\overset{\mathrm{(A5)}}{\leq} & \frac{1}{v_{k}}\lambda_{\max}\left(\mathbf{Q}_{\mathbf{x},k}\right)+\frac{4\mathbf{x}^{H}\mathbf{Q}_{\mathbf{x},k}\mathbf{x}}{v_{k}^{3}}\mathbf{x}^{H}\overline{\mathbf{Q}}_{\mathbf{x},k}\overline{\mathbf{Q}}_{\mathbf{x},k}\mathbf{x}.\label{eq:H-3}
\end{align}
Equation (A5) in (\ref{eq:LAMDA-1}) is due to the fact that $\overline{\mathbf{Q}}_{\mathbf{x},k}$
is a rank-1 positive semidefinite metrix and $-\mathrm{Re}\left(\mathbf{x}^{\mathrm{H}}\mathbf{Q}_{\mathbf{x},k}\overline{\mathbf{Q}}_{\mathbf{x},k}\mathbf{x}\right)$
is non-positive. Take (\ref{eq:H-2}) and (\ref{eq:H-3}) into (\ref{eq:H-1}),
we obtain

\begin{align}
 & \lambda_{\max}(\mathcal{H}_{u_{k}})\nonumber \\
\leq & \frac{1}{\omega_{k}^{2}\epsilon}\left(\frac{\mathbf{x}^{H}\mathbf{Q}_{\mathbf{x},k}\mathbf{Q}_{\mathbf{x},k}\mathbf{x}}{v_{k}^{2}}-4\mathrm{Re}\left(\frac{\mathbf{x}^{H}\mathbf{Q}_{\mathbf{x},k}\mathbf{x}\mathbf{x}^{H}\mathbf{Q}_{\mathbf{x},k}\overline{\mathbf{Q}}_{\mathbf{x},k}\mathbf{x}}{v_{k}^{3}}\right)+\frac{6\left(\mathbf{x}^{H}\mathbf{Q}_{\mathbf{x},k}\mathbf{x}\right)^{2}}{v_{k}^{4}}\mathbf{x}^{H}\mathbf{\overline{Q}}_{\mathbf{x},k}\overline{\mathbf{Q}}_{\mathbf{x},k}\mathbf{x}\right)\\
 & +\frac{1}{\omega_{k}^{2}\epsilon}\frac{\mathbf{x}^{H}\mathbf{Q}_{\mathbf{x},k}\mathbf{x}}{v_{k}^{2}}\lambda_{\max}\left(\mathbf{Q}_{\mathbf{x},k}\right)+\frac{1}{\epsilon}\lambda_{\max}(\mathcal{H}_{u_{k}})\nonumber \\
\overset{\mathrm{(A6)}}{\leq} & \begin{cases}
\frac{1}{\omega_{k}^{2}\epsilon\sigma_{k}^{4}}\left((2+5\omega_{k})a-\frac{4ab}{\sigma_{k}^{2}}+\frac{6ab^{2}}{\sigma_{k}^{4}}\right) & \textrm{if }\mathbf{x}=\mathbf{e}\\
\frac{b^{2}}{\omega_{k}^{2}\epsilon\sigma_{k}^{4}NN_{RF}}\left(2+\frac{(5\omega_{k}-4b)}{\sigma_{k}^{2}}+\frac{6b^{2}}{\sigma_{k}^{4}}\right) & \textrm{\textrm{if }}\mathbf{x}=\mathbf{\mathrm{vec}({\bf A})}\\
\frac{NN_{RF}b^{2}}{\omega_{k}^{2}\epsilon\sigma_{k}^{4}P_{max}}\left(2+\omega_{k}+\frac{6b^{2}}{\sigma_{k}^{4}}\right) & \textrm{if }\mathbf{x}=\mathrm{vec}({\bf D}).
\end{cases}\label{eq:LEMA-2}
\end{align}
The inequality (A6) in (\ref{eq:LEMA-2}) is obtained by using the
similar violent relaxations as in (\ref{eq:LAMDA-1}). The results
of the relaxations are straightforwardly given below 
\begin{align*}
\mathbf{x}^{\mathrm{H}}\mathbf{Q}_{\mathbf{x},k}\mathbf{x} & \leq b,\\
\mathbf{e}^{H}\mathbf{Q}_{\mathbf{e},k}\mathbf{Q}_{\mathbf{e},k}{\bf e} & \leq a,\\
\mathbf{e}^{H}\overline{\mathbf{Q}}_{\mathbf{e},k}\overline{\mathbf{Q}}_{\mathbf{e},k}{\bf e} & \leq a,\\
\mathrm{Re}\left(\mathbf{e}^{\mathrm{H}}\mathbf{Q}_{\mathbf{e},k}\overline{\mathbf{Q}}_{\mathbf{e},k}\mathbf{e}\right) & \leq a,\\
\mathrm{vec}({\bf A})^{\mathrm{H}}\mathbf{Q}_{\mathbf{A},k}\mathbf{Q}_{\mathbf{A},k}\mathrm{vec}({\bf A}) & \leq\frac{b^{2}}{NN_{RF}},\\
\mathrm{vec}({\bf A})^{\mathrm{H}}\overline{\mathbf{Q}}_{\mathbf{A},k}\overline{\mathbf{Q}}_{\mathbf{A},k}\mathrm{vec}({\bf A}) & \leq\frac{b^{2}}{NN_{RF}},\\
\mathrm{Re}\left(\mathrm{vec}({\bf A})^{\mathrm{H}}\mathbf{Q}_{\mathbf{A},k}\overline{\mathbf{Q}}_{\mathbf{A},k}\mathrm{vec}({\bf A})\right) & \leq\frac{b^{2}}{NN_{RF}},\\
\mathrm{vec}({\bf D})^{\mathrm{H}}\mathbf{Q}_{\mathbf{D},k}\mathbf{Q}_{\mathbf{D},k}\mathrm{vec}({\bf D}) & \leq\frac{b^{2}NN_{RF}}{P_{max}},\\
\mathrm{vec}({\bf D})^{\mathrm{H}}\overline{\mathbf{Q}}_{\mathbf{D},k}\overline{\mathbf{Q}}_{\mathbf{D},k}\mathrm{vec}({\bf D}) & \leq\frac{b^{2}NN_{RF}}{P_{max}},\\
\mathrm{Re}\left(\mathrm{vec}({\bf D})^{\mathrm{H}}\mathbf{Q}_{\mathbf{D},k}\overline{\mathbf{Q}}_{\mathbf{D},k}\mathrm{vec}({\bf D})\right) & =0,\\
\lambda_{\max}\left(\mathbf{Q}_{\mathbf{e},k}\right) & \leq\frac{b}{UM+1},\\
\lambda_{\max}\left(\mathbf{Q}_{\mathbf{A},k}\right) & \leq\frac{b}{NN_{RF}},\\
\lambda_{\max}\left(\mathbf{Q}_{\mathbf{D},k}\right) & \leq\frac{bNN_{RF}}{P_{max}}.
\end{align*}

The proof is completed.

 \bibliographystyle{IEEEtran}
\bibliography{bibfile}

\end{document}